\documentclass{article}
\usepackage{graphicx} 
\usepackage{amsthm, amsmath, amsfonts,mathtools, fullpage, graphicx, wrapfig, tikz, caption, subcaption}
\usepackage{natbib}
\usepackage{amssymb}
\usepackage{verbatim}
\usepackage{dsfont}
\usepackage{relsize}
\usepackage{listings}
\usepackage{xcolor}
\usepackage{hyperref}
\usepackage{graphicx}
\usepackage{multirow}
\usepackage[linesnumbered,ruled,vlined]{algorithm2e}

\newtheorem{theorem}{Theorem}
\newtheorem{lemma}{Lemma}
\newtheorem{corollary}{Corollary}

\newtheorem{remark}{Remark}
\newcommand{\norm}[1]{\left\lVert#1\right\rVert}
\newtheorem{assumption}{Assumption}

\usepackage{amsmath,amssymb,xcolor}
\usepackage{pgf,tikz}
\usetikzlibrary{arrows,shapes.arrows,shapes.geometric,shapes.multipart,decorations.pathmorphing,positioning,shapes.swigs,}

\usepackage{setspace}

 \usepackage[vmargin=1.18in,hmargin=1.1in]{geometry}

\title{Difference-in-Differences for Heterogeneous Effects of General Treatment Strategies Accounting for Time-Varying Covariates}
\author{Oliver Dukes$^{1}$\footnote{equal contribution}, Zach Shahn$^{2*}$, James Robins$^{3}$, and Andrea Rotnitzky$^{4}$\\
$^1$: Ghent University, Ghent,  Belgium\\
$^2$: CUNY Graduate School of Public Health and Health Policy,  New York,  NY,  USA\\
$^3$: Harvard T. H. Chan School of Public Health,  Boston,  MA,  USA\\
$^4$: University of Washington, Seattle, WA, USA}

\date{\today}

\begin{document}

\maketitle

\begin{abstract}
We introduce a class of marginal Structural Nested Mean Models (mSNMMs) that enable modeling time-varying effect modification by a subset of covariates required as adjustment variables. We focus on identifiability and estimation under conditional parallel trends assumptions, so mSNMMs are applicable in settings where time-varying Difference-in-Differences (DiD) methods are typically applied. Modeling effect modification by lower dimensional covariates reduces the risk of model misspecification and can improve interpretability relative to previously developed `full SNMMs'. mSNMMs enjoy two key advantages over alternative time-varying DiD methods. First, the parallel trends assumptions that they require are more plausible in the presence of time-varying confounding. Second, they are able to model effects of \textit{sustained} interventions even when treatment switches on and off in the data. We illustrate the methodology by applying mSNMMs to estimate how the effects of Medicaid expansion (a time-varying treatment) on local uninsurance rates vary with unemployment rate prior to expansion (a time-varying covariate), adjusting for additional time-varying demographic and economic variables.

\end{abstract}

\section{Introduction}
When the goal of a causal analysis is to learn about effect modification, there are multiple roles covariates might play. They might be effect modifiers of interest, adjustment variables to reduce bias and/or improve precision, or a combination. In this paper, we consider the scenario in which an investigator must adjust for a large set of covariates but, for reasons of interpretability or practicality, is only interested in effect modification by a subset (or other function) of those covariates. We focus in particular on settings in which the investigator is willing to make a conditional (given the set of adjustment variables) parallel trends assumption of the sort that underlies difference-in-differences (DiD) analyses. For example, suppose interest centers on how the effect of a minimum wage hike on county level unemployment rate varies with poverty rate prior to the hike. Parallel trends may only hold conditional on various demographic and economic county level characteristics in addition to poverty rate, but a researcher might nonetheless like to avoid modeling effect heterogeneity as a function of all of these additional covariates. 


Structural nested mean models (SNMMs) are models of (possibly time-varying) effect heterogeneity as a function of covariates \citep{jamie1994,jamie1997,jamie2004, snmm_review2014}. SNMMs were originally shown to be identified under a sequential exchangeability or no unobserved confounding assumption. Recently, \citet{us2022} showed that SNMMs are also identified under a conditional parallel trends assumption. However, \citet{us2022} only considered the case where the SNMM models effect modification by the full set of covariates that must be conditioned on to satisfy the parallel trends assumption. If there are many covariates, severe misspecification of this model may become likely. Further, even if correctly specified, a high dimensional SNMM may be difficult to interpret. 

Thus, marginal SNMMs (mSNMMs) that only model effect modification by a low dimensional function (usually a subset) of the covariates are desirable both to reduce bias from model misspecification and for their comparative interpretability. These models were introduced in chapter 6 of \citet{laan2003unified} in the separate context of generalizing  Robins' structural nested failure time model, which is a model for the treatment effect on the entire \textit{distribution} of an outcome, rather than just the mean. Under sequential conditional exchangeability, \citet{jamie2004} describes inverse probability of treatment weighting estimators for mSNMMs (see their Section 7.3).  We show that mSNMMs are identified under conditional parallel trends assumptions given the full set of covariates and provide doubly robust, semiparametric efficient estimators. In the point exposure setting, estimators closely related to ours were proposed in the PhD thesis of \citet{lei2011improved} under a conditional exchangeability assumption. Our class of estimators includes the inverse probability of treatment weighting estimator of \citet{abadie2005} as a special case. In the longitudinal setting, our identification and semiparametric estimation results are new and do not follow from the aforementioned references.

In concurrent work developed independently, \citet{lan2025meta} obtain results that partially overlap with our own. In the point exposure setting, they develop nonparametric estimators of the conditional treatment effect in the treated (marginal over additional adjustment covariates) under parallel trends, based on the same doubly robust conditional moment restriction as we use. However, in the longitudinal setting, their results deviate from ours. They align instead with other recent DiD methods for estimating time-varying effects \citep{roth2023,chaisemartin_review} in two important ways: they make similar parallel trends assumptions, and they consider the same limited range of `staggered adoption' treatment effects. We will argue that mSNMMs have advantages on both of these fronts.


First, a source of consternation in the time-varying DiD literature has been estimation of effects of sustained interventions when, in the observed data, treatment might switch on and off; see Section 3.4 of \citet{roth2023} and \citet{CandDH2024difference}. It is possible to identify these effects via either full SNMMs \citep{us2022} or mSNMMs under particular parallel trends assumptions appropriate for dynamic treatment regimes (see \citet{renson2023identifying} for an alternative approach). Due to model misspecification concerns regarding full SNMMs, investigators might prefer mSNMMs for this task. 
Alternative DiD approaches settle for estimating the effect of first initiating treatment, marginalizing over future treatment patterns under the observational regime. 

Secondly, in contrast with the parallel trends assumptions underlying alternative time-varying DiD approaches, our parallel trends assumption \citep{us2022} might plausibly hold even in the presence of observed time-varying confounders. With the exception of \citet{renson2023identifying}, alternative methods can only plausibly adjust for time-invariant baseline covariates. The comparison between our parallel trends assumption and the `never treated' parallel trends assumption of \citet{callaway2021difference} is discussed at length in Section 8.2 of \citet{us2022} and in Remark \ref{remark_pt} below. Briefly, the issue is that their parallel trends assumptions are made with respect to a control group defined in terms of \textit{observed} future treatment status, e.g. the `never treated' \citet{callaway2021difference}, which would be influenced by future values of the confounders. Our parallel trends assumption is that at each time $t$, future expected \textit{counterfactual} untreated outcome trends are equal in units treated or untreated at $t$ (conditional on covariate and treatment history up to $t$). Because our parallel trends assumption contrasts comparable groups, it might hold even if time-varying covariates influence treatment and trends. 

Outside of \citet{us2022} and \citet{lan2025meta}, existing time-varying DiD methods do not enable modeling of effect modification by time-varying covariates, and are therefore not directly comparable to mSNMMs when effect heterogeneity is of interest. However, we note that if an investigator is not interested in effect heterogeneity but prefers our parallel trends assumption to alternatives or wishes to estimate effects of sustained interventions, they can employ mSNMMs in which the set of effect modifiers of interest is specified to be the empty set. In staggered adoption settings, these can be used to target the same estimands as other approaches, i.e. what \citet{callaway2021difference} call `$ATT(g,t)$'.


The organization of the paper is as follows. In Section 2, we discuss mSNMMs for the point exposure setting and provide semiparametric efficient estimators. In Section 3, we provide identification results and estimators for time-varying mSNMMs. In Section 4, we discuss mSNMMs for dynamic treatment regimes in which treatment assignment might depend on evolving treatment and covariate history. We emphasize the special case of the `empty set' mSNMM for estimating effects of sustained treatment when treatment switches on and off in the observed data. In Section 5, we present simulation results. In Section 6, we apply mSNMMs to estimate heterogeneous effects of Medicaid expansion on uninsurance rates as a function of unemployment rate. In Section 7, we conclude.

\section{The Point Exposure Setting}
\subsection{Identification}

For didactic reasons, in this section we discuss identification and estimation in the point-exposure case. Suppose we observe a cohort of $N$ units indexed by $i \in \{1,\ldots,N\}$ and collect data on covariates $L_0$, pre-treatment outcome measurement $Y_0$, treatment $A$, and post-treatment outcome $Y_1$. We assume the temporal ordering $O=(L_0,Y_0,A,Y_1)$. Let $\tilde{L}_0$ denote some function (e.g. a subset) of $L_0$.  We will let $\mathbb{P}_N$ denote the empirical measure and $\mathbb{P}$ denote the true data generating law. The expectation $\mathbb{E}_P[\cdot]$ is evaluated at a law $P$; if not stated otherwise, the expectation $\mathbb{E}[\cdot]$ is evaluated at the true law $\mathbb{P}$. Finally, for some function $g(O)$, we let $\norm{g}$ denote $\{\int g(O)^2 d\mathbb{P}(O)\}^{1/2}$; we use $\norm{\cdot}_2$ to denote the Euclidean norm. 

In what follows, we are interested in inferring the conditional effect of treatment on the treated given $\tilde{L}_0$, i.e.
\begin{equation}\label{pe_msnmm}
\gamma(\tilde{l}_0)\equiv \mathbb{E}\{Y_1(1)-Y_1(0)|\tilde{L}_0=\tilde{l}_0,A=1\},
\end{equation}
where we adopt the usual counterfactual notation $Y_1(a)$ for the value of $Y_1$ had, possibly contrary to fact, treatment $A$ been set to value $a$. We will make the parallel trends assumption conditional on the full covariate $L_0$
\citep{abadie2005}.
\begin{assumption}\label{pe_pt}{(Conditional parallel trends)}
    \begin{align*}
\mathbb{E}\{Y_1(0)-Y_0|L_0,A=1\}=\mathbb{E}\{Y_1(0)-Y_0|L_0,A=0\}
\end{align*}
with probability one. 
\end{assumption}
We further make the following two assumptions that are standard in causal inference.
\begin{assumption}\label{consistency_pe}{(Consistency)}
\begin{align*}
   Y_1(a)=Y_1 \text{ whenever }A=a.
\end{align*}
\end{assumption}
\begin{assumption}\label{positivity_pe}{(Positivity)}
\begin{align*}
f_{A|L_0}(0|l_0)>0 \text{ whenever } f_{L_0}(l_0)>0.
\end{align*}
\end{assumption}
The consistency assumption states that the counterfactual outcome corresponding to the factual treatment is equal to the observed outcome. 

The following theorem also follows from \citet{lan2025meta}.
\begin{theorem}\label{pe_id}
    Under Assumptions \ref{pe_pt}-\ref{positivity_pe}, $\gamma(\tilde{l}_0)$ is identified as\\ $\mathbb{E}\{Y_1-Y_0-\mathbb{E}(Y_1-Y_0|L_0,A=0)|\tilde{L}_0=\tilde{l}_0,A=1\}$.
\end{theorem}
\begin{proof}
    \begin{align*}
    {\gamma}(\tilde{l}_0)&= \mathbb{E}\{Y_1(1)-Y_1(0)|\tilde{L}_0=\tilde{l}_0,A=1\}\\
    &=\mathbb{E}\{Y_1(1)-Y_0|\tilde{L}_0=\tilde{l}_0,A=1\} - \mathbb{E}\{Y_1(0)-Y_0|\tilde{L}_0=\tilde{l}_0,A=1\}\\
    &=\mathbb{E}(Y_1-Y_0|\tilde{L}_0=\tilde{l}_0,A=1) - \mathbb{E}[\mathbb{E}\{Y_1(0)-Y_0|L_0,A=1\}|\tilde{L}_0 =\tilde{l}_0,A=1]\\
    &=\mathbb{E}(Y_1-Y_0|\tilde{L}_0 =\tilde{l}_0,A=1) - \mathbb{E}[\mathbb{E}\{Y_1(0)-Y_0|L_0,A=0\}|\tilde{L}_0=\tilde{l}_0,A=1]\\
    &=\mathbb{E}(Y_1-Y_0|\tilde{L}_0 =\tilde{l}_0,A=1)-\mathbb{E}\{\mathbb{E}(Y_1-Y_0|L_0,A=0)|\tilde{L}_0=\tilde{l}_0,A=1\}.
\end{align*}
The first equality is the definition of $\gamma(\tilde{l}_0)$. The second simply adds and subtracts $\mathbb{E}(Y_0|\tilde{L}_0=\tilde{l}_0,A=1)$. The third follows from consistency and nested expectation. The fourth invokes parallel trends (\ref{pe_pt}). The final equality follows from consistency and has eliminated all counterfactuals, establishing identification.
\end{proof}

\subsection{Semiparametric estimation}


We now turn our attention to estimation of a parametric single time point mSNMM (or just mSMM, standing for `marginal structural mean model', as there is no `nesting' in the point exposure setting). A parametric mSMM assumes that for each possible value $\tilde{l}_0$ of $\tilde{L}_0$, there exists a known function $\gamma(\tilde{l}_0;\cdot):\Psi \mapsto \mathbb{R}$, $0\in \Psi \subset \mathbb{R}^p$, satisfying $\gamma(\tilde{l}_0;0)=0$, such that 
\begin{equation}\label{smm}
\gamma(\tilde{l}_0)=\gamma(\tilde{l}_0;\psi^*)
\end{equation}
where $\psi^*\in \Psi$ is unknown. We call (\ref{smm}) a \textit{marginal} SMM because it models effect heterogeneity given $\tilde{L}_0$ marginal over other information contained in $L_0$. The identification assumptions of Theorem \ref{pe_id} can be shown not to restrict the distribution of the observed data $O$. Furthermore, under these assumptions, $\gamma(\tilde{l}_0)=\mathbb{E}\{Y_1-Y_0-\mathbb{E}(Y_1-Y_0|L_0,A=0)|\tilde{L}_0=\tilde{l}_0,A=1\}$. Thus, under (\ref{smm}), our statistical model $\mathcal{M}$ comprises the collection of distributions $P$ for which the conditional expectation function $\mathbb{E}_{P}\{Y_1-Y_0-\mathbb{E}_{P}(Y_1-Y_0|L_0,A=0)|\tilde{L}_0,A=1\}$ is equal to $\gamma(\tilde{L}_0;\psi)$ for some $\psi \in \Psi$; we assume $\gamma(\cdot;\cdot)$ is smooth in $\psi$. 


In this section, we will characterize the set of influence functions of all regular and asymptotically linear (RAL) estimators of $\psi^*$ under model $\mathcal{M}$.  This will then lead to the construction of doubly-robust machine learning (DR-ML) estimators of $\psi^*$. Let $\pi^*(L_0)\equiv \mathbb{E}(A|L_0)$ and $\mu^*(L_0)\equiv\mathbb{E}(Y_1-Y_0|L_0,A=0)$. 
\begin{theorem}\label{theorem:pe_IFs}
(i) Under the model $\mathcal{M}$, the set of influence functions for all RAL estimators of $\psi^*$ is given by
\begin{equation*}
    \left\{D(\psi^*,q)^{-1}\phi(O;\psi^*,q,\pi^*,\mu^*): q\textrm{ unrestricted} \right\}\cap \mathcal{L}_2
\end{equation*}
    where for any $\psi\in\Psi$ and any arbitrary functions $\pi(L_0)$ and $\mu(L_0)$ (such that $\pi(L_0)\neq1$ with probability one),
\begin{align*}
 D_P(\psi,q)\equiv& \frac{\partial }{\partial \psi}\mathbb{E}_P\left\{q(\tilde{L}_0)A\gamma(\tilde{L}_0;\psi)\right\}|_{\psi=\psi},\\
\phi(O;\psi,q,\pi,\mu)\equiv& q(\tilde{L}_0)\left[\frac{A-\pi(L_0)}{1-\pi(L_0)}\{Y_1-Y_0-\mu(L_0)\}-A\gamma(\tilde{L}_0;\psi)\right].
\end{align*}
Here, $q(\tilde{L}_0)$ is an arbitrary $p$-dimensional function of $\tilde{L}_0$ such that $ D(\psi^*,q)^{-1}$ exists and $\mathcal{L}_2\equiv \mathcal{L}_2(\mathbb{P})$ is the Hilbert space of real-valued mean-zero functions of $O$ with finite variance under $\mathbb{P}$. \\
(ii) Let $q_{eff}(\tilde{L}_0)$ denote the optimal index function, in the sense that under model $\mathcal{M}$, no RAL estimator of $\psi^*$ can have smaller asymptotic variance (in a positive definite sense) than one that uses $q_{eff}(\tilde{L}_0)$. Then
\[q_{eff}(\tilde{L}_0)=J(\tilde{L}_0;\psi^*)^T \textrm{var}\left[\frac{A-\pi^*(L_0)}{1-\pi^*(L_0)}\{Y_1-Y_0-\mu^*(L_0)\}-A\gamma(\tilde{L}_0;\psi^*)|\tilde{L}_0\right]^{-1}\]
where, $J(\tilde{L}_0;\psi^*)=\mathbb{E}(A|\tilde{L}_0)\frac{\partial}{\partial\psi}\gamma(\tilde{L}_0;\psi)|_{\psi=\psi^*}$. \\
(iii) The estimating function $\phi(O;\psi^*,q,\pi^\dagger,\mu^\dagger)$ is doubly robust. That is, let $\pi^\dagger(L_0)$ and $\mu^\dagger(L_0)$ be arbitrary functions of $L_0$. We have, for any $p$-dimensional function $q(\tilde{L}_0)$,
\begin{equation}\label{pe_est_eq}
     \mathbb{E}\left\{\phi(O;\psi^*,q,\pi^\dagger,\mu^\dagger)\right\}=0
\end{equation}
 if either $\mu^\dagger=\mu^*$ or $\pi^\dagger=\pi^*$.
\end{theorem}
A proof of this result and all others can be found in the Appendix. Under the assumption of conditional exchangeability,  closely related estimators were derived in \citet{lei2011improved}.

We will now use Theorem \ref{theorem:pe_IFs} to motivate specific estimators for $\psi^*$. To construct estimators in practice, we need to estimate the unknown nuisance functions $\mu$ and $\pi$. We consider doubly robust machine learning (DR-ML) estimators combined with cross-fitting \citep{schick1986asymptotically,van2000asymptotic,chernozhukov2018double},  in which the nuisance functions are estimated by nonparametric estimators/machine learning algorithms. Algorithm \ref{algorithm:cfdrml} describes the implementation of the cross-fit estimator $\hat{\psi}^{cf}\equiv \hat{\psi}^{cf}(q,\hat{\pi},\hat{\mu})$ where $q(L_0)$ is again a user chosen vector function sharing the dimension of $\psi^*$. Here and elsewhere, the number of folds $S$ is assumed to be a fixed positive integer greater than or equal to 2. 

\begin{algorithm}
\caption{Implementation of a cross-fit DR-ML estimator: point-exposure setting}\label{algorithm:cfdrml}
 Randomly split the $N$ units into $S$ disjoint folds of size $n_s\approx N/S$.\\
For each $s=1,...,S$, construct estimators $\hat{\mu}^{(-s)}$ of $\mu^*$ and $\hat{\pi}^{(-s)}$ of $\pi^*$, using all the units except those in fold $s$.\\
For each $s=1,...,S$, compute $\hat{\psi}^{(s)}$ from the $s$-th fold as the (assumed unique) solution to the estimating equations $\mathbb{P}_{n_s}\{\phi(O;\psi,q,\hat{\pi}^{(-s)},\hat{\mu}^{(-s)})\}=0$ where $\mathbb{P}_{n_s}$ refers here to the empirical measure in the $s$-th fold.\\
Take the weighted average of the $S$ estimators:  $\hat{\psi}^{cf}=\sum^{S}_{s=1}\left(\frac{n_s}{N}\right)\hat{\psi}^{(s)}$.
\end{algorithm}

The following theorem establishes the asymptotic properties of $\hat{\psi}^{CF}$. 
\begin{theorem}\label{estimation_theorem}
Suppose, in addition to Assumption \ref{positivity_pe}, that the following conditions hold:
\begin{enumerate}
    \item The map $\psi \mapsto \mathbb{P}\{\phi(o;\psi,q,\pi,\mu)\}$ is differentiable at $\psi^*$ uniformly in $(\pi,\mu)$. 
    \item The matrix $D(\psi^*,q)$ is nonsingular.
    \item $\{1-\hat{\pi}^{(-s)}(L_0)\}\geq c>0$ and $\{1-\pi^*(L_0)\}\geq c>0$  with probability one for some constant $c$ and $s=1,...,S$.
    \item For a chosen $p$-dimensional function $q(\tilde{L}_0)$, the maximum absolute value of any element in $q(\tilde{L}_0)$ is upper bounded by a universal constant $C_1$ with probability one.
    \item $\mathbb{E}[\{Y_1-Y_0-\mu^*(L_0)-A\gamma(\tilde{L}_0;\psi^*)\}^2|L_0,A]$ is upper bounded by $C_2<\infty$ with probability one.
    \item $\hat{\psi}^{(s)}-\psi^*=o_{\mathbb{P}}(1)$, $\norm{\hat{\pi}^{(-s)} - \pi^*}=o_\mathbb{P}(1)$ and $\norm{\hat{\mu}^{(-s)}-\mu^*}=o_\mathbb{P}(1)$ for $s=1,...,S$.
    \item $\norm{\hat{\pi}^{(-s)} - \pi^*}\norm{\hat{\mu}^{(-s)}-\mu^*}=o_\mathbb{P}(N^{-1/2})$ for $s=1,...,S$.
    \item For any fixed ($\pi,\mu$), the class $\left\{\phi(o;\psi,q,\pi,\mu):\psi\in \Psi\right\}$ is Donsker in $\psi$. 
\end{enumerate}
Then $\hat{\psi}^{cf}$ satisfies the expansion
\begin{equation*}
\sqrt{N}(\hat{\psi}^{cf}-\psi^{*})=D(\psi^*,q)^{-1} \frac{1}{\sqrt{N}}\sum_{i=1}^N \phi(O;\psi^*,q,\pi^*,\mu^*) +o_\mathbb{P}(1)
\end{equation*}
In particular, $\sqrt{N}(\hat{\psi}^{cf}-\psi^{*})$ converges to a mean zero normally distributed variable with variance equal to $\textrm{var}\{D(\psi^*,q)^{-1}\phi(O;\psi^*,q,\pi^*,\mu^*)\}$, where  $\textrm{var}\{D(\psi^*,q)^{-1}\phi(O;\psi^*,q,\pi^*,\mu^*)\}$ is minimized (in a positive definite sense) at $q$ equal to the function $q_{eff}$ defined in Theorem \ref{theorem:pe_IFs}(ii). 
\end{theorem}
This theorem is a special case of the more general result for time-varying exposures, given in the following section. An asymptotically efficient estimator of $\psi^*$ can then be constructed via nonparametric estimation of $q_{eff}(\tilde{L}_0)$. A condition for asymptotic linearity for $\hat{\psi}^{cf}$ is therefore that the product of the rates of convergence of the nuisance estimators for $\mu^*(L_0)$ and $\pi^*(L_0)$ is $o_\mathbb{P}(n^{-1/2})$. This property has been referred to as `rate double robustness' \citep{smucler2019unifying}. The remaining conditions are standard regularity conditions for $Z$-estimators; see e.g. Chapter 5 of \citet{van2000asymptotic}. Here and elsewhere, we propose to estimate the variance of $\hat{\psi}^{cf}$ using a sandwich estimator: in this case, this is
\[\hat{D}(\hat{\psi}^{cf},q)^{-1}\left[\frac{1}{S}\sum^S_{s=1}\mathbb{P}_{n_s}\{\phi(O;\hat{\psi}^{cf},q,\hat{\pi}^{(-s)},\hat{\mu}^{(-s)})\phi(O;\hat{\psi}^{cf},q,\hat{\pi}^{(-s)},\hat{\mu}^{(-s)})^T\}\right]\{\hat{D}(\hat{\psi}^{cf},q)^{-1}\}^T/N\]
where 
\[\hat{D}(\hat{\psi}^{cf},q)=\frac{1}{S}\sum^{S}_{s=1} \mathbb{P}_{n_s}\left\{q(\tilde{L}_0)A\frac{\partial }{\partial \psi}\gamma(\tilde{L}_0;\psi)|_{\psi=\hat{\psi}^{cf}}\right\};\]
see Theorem 3.2 of \citep{chernozhukov2018double} for a justification.

Analysts can also choose to estimate $\psi^*$ by using parametric working models $\pi(L_0;\alpha)$ for $\pi^*(L_0)$ and $\mu(L_0;\beta)$ for $\mu^*(L_0)$, where for each possible value $\tilde{l}_0$ of $\tilde{L}_0$, 
$\alpha \mapsto \pi(l_0;\alpha)$ and $\beta  \mapsto \mu(l_0;\beta)$ are known functions that are smooth in finite-dimensional parameters $\alpha$ and $\beta$ respectively. In this case, cross-fitting is unnecessary. Estimators of $\psi^*$ constructed in this way will be consistent and asymptotically normal if either the propensity score or the outcome regression model is correctly specified. We also note that such an estimator reduces to the doubly robust estimator of the average treatment effect in the treated (ATT) proposed by \citet{sant2020doubly} in the special case that $\tilde{L}_0$ is the empty set. 




\begin{remark}\label{remark_abadie}
Suppose that the mSMM is incorrectly specified; then for every fixed $q$, our estimator will converge to a functional that is the solution of the  population estimating equation:
\[0=\mathbb{E}[q(\tilde{L}_0)\{Y_1-Y_0-\mu^*(L_0)-A\gamma(\tilde{L}_0;\psi^*)\} ].\]
In particular, if we choose $q(\tilde{L}_0)=\frac{\partial}{\partial\psi^T}\gamma(\tilde{L}_0;\psi)|_{\psi=\psi^*}$, then the solution coincides with the least squares projections of the conditional average treatment effect in the treated onto the low dimensional working model $\gamma(\tilde{L}_0;\psi^*)$. \citet{abadie2005} proposes an inverse probability weighted-estimator of this least squares projection parameter, using a nonparametric logistic series estimator of the propensity score. By taking a doubly robust approach, our estimators more easily facilitate valid inference using data-adaptive nuisance estimation (outside of series methods). Furthermore, although \citet{abadie2005} did not consider the mSMM parameter as the functional of inference, if one assumes that the mSMM is correctly specified, then our Theorem \ref{theorem:pe_IFs} shows that their estimator is inefficient. 

\end{remark}

 \begin{remark}
     We derived the set of influence functions (and therefore the class of RAL estimators) under model $\mathcal{M}$ using pathwise derivative calculations. However, we describe in Appendix \ref{section_appendix_alternate} an alternative derivation of our estimation results, which may be more intuitive to those familiar with \textit{marginal structural models} in \citet{robins2000marginal}. It also facilitates a connection with the estimators derived in \citet{us2022} for SNMMs under parallel trends. 

\end{remark}
\citet{lan2025meta} consider meta-learners of the conditional treatment effect in the treated under parallel trends, developing a generic loss minimization approach under a function class that enjoys double robustness properties. Had they restricted their function class to a parametric class (potentially with dimension growing with sample size), then it would deliver estimators with the same properties as in Theorem \ref{estimation_theorem}.



\section{The time-varying setting}\label{section_time_varying}
We now focus on deriving methodology for estimating heterogeneous effects of a time-varying intervention as a function of the history of variables of interest in the currently treated units. For example, the minimum wage might be increased at different times in different counties, and a researcher might wish to study how the effects of wage increases on unemployment vary with poverty rate trajectory leading up to the wage increase and additionally wish to adjust for county demographic variables. We formalize the problem as follows.   

Assume that each unit is observed at regular intervals from baseline
time $0$ through end of followup time $K$, and there is no loss to
follow-up. At each time $m=0,\ldots,K-1$, data on $(Z_m,Y_m,A_m)$ are available, and at time $K$ we observe $Y_K$. $Z_m$ and $Y_m$ are realized prior to $A_m$ and subsequent to $A_{m-1}$. Then $A_m$ denotes the treatment received at time $m$, $Y_m$ denotes the outcome of interest at time $m$, and $Z_m$ denotes a vector of covariates at time $m$ excluding $Y_m$. Hence, $Z_0$
constitutes the vector of baseline covariates other than $Y_0$. Throughout, for an arbitrary vector $X=(X_0,\ldots,X_K)$ and $0\leq m\leq K$, we let $\bar{X}_m\equiv(X_0,\ldots,X_m)
$ and $\underline{X}%
_m\equiv(X_m,\ldots,X_K)$. Whenever $X_{-1}$ appears, we define it to take the null value with probability 1. Define $\bar{L}_m$ to be a function of $(\bar{Z}_m,\bar{Y}_{m})$, i.e. a possibly high dimensional summary of the covariate and outcome history through time $m$. We can regard $\bar L_m$ as analyst chosen. All results are stated in terms of an arbitrary $\bar L_m$, but it should be understood that assumptions might hold for some choices and not others depending on the data generating process. In particular, Remark 6 and Section 8.2 of \citet{us2022} discuss how it is in general undesirable to include outcome level histories $\bar Y_m$ directly in $\bar L_m$ (especially the immediately prior outcome $Y_m$) because it leads to assumption violations. However, increment histories $\bar\Delta_m$ where $\Delta_j\equiv Y_j-Y_{j-1}$ do not cause similar problems. Let $\tilde{\bar{L}}_m$ denote some function (e.g. a subvector) of $\bar{L}_m$ such that $\tilde{\bar{L}}_m$ is increasing in $m$, with `increasing' meaning that $\tilde{\bar{L}}_m$ is a function of $\tilde{\bar{L}}_{m+1}$ for all $m$. For example, for $\bar{L}_m=(\bar{L}_{m1},\ldots,\bar{L}_{mp}$), $\tilde{\bar{L}}_m$ might be $(\bar{L}_{m2},\bar{L}_{m5})$. We adopt the counterfactual framework for time-varying treatments
\citep{robins1986new} which posits that corresponding to each time-varying
treatment regime $\bar{a}_m$, each unit has a counterfactual or potential
outcome $Y_{m+1}(\bar{a}_{m})$ that would have been observed had that
unit received treatment regime $\bar{a}_m$. 

Now consider the so-called `blip' functions
\begin{equation}\label{blip_models}
    \gamma_{mk}(\tilde{\bar{l}}_m,\bar{a}_{m-1})\equiv \mathbb{E}\{Y_k(\bar{a}_{m-1},1,\underline{0})-Y_k(\bar{a}_{m-1},\underline{0})|\tilde{\bar{L}}_m=\tilde{\bar{l}}_m,\bar{A}_{m-1}=\bar{a}_{m-1},A_m=1\}.
\end{equation}
 with $\gamma_{mk}(\tilde{\bar{l}}_m,\bar{a}_{m-1})$ defined to be $0$ whenever $Pr(A_m=1|\tilde{\bar{L}}_m=\tilde{\bar{l}}_m,\bar{A}_{m-1}=\bar{a}_{m-1})=0$. We call (\ref{blip_models}) blip functions because they represent the conditional effect (in the treated at time $m$) of one final blip of treatment at time $m$ followed by no further treatment compared to no treatment at time $m$ or thereafter given (selected) covariate and treatment history. $\gamma_{mk}(\tilde{\bar{l}}_m,\bar{a}_{m-1})$ thus characterizes time-varying effect heterogeneity of blips of treatment. For example, (\ref{blip_models}) might represent how the lasting effects of a single year of small class size in year $m$ ($A_m=1)$ on subsequent test scores ($Y_k$, $k>m)$ depend on students' history of socioeconomic status and class size ($\tilde{\bar{L}}_m,\bar{A}_{m-1}$). Recalling that we defined $X_{-1}$ to be null, note that when $K=1$, $\gamma_{01}(\tilde{l}_0)$ corresponds to the $\gamma(\tilde{l}_0)$ from earlier sections. 


\subsection{Nonparametric Identification}
The following assumption formalizes the notion of parallel trends conditional on the full history of adjustment variables $\bar{L}_m$ \citep{us2022}:
\begin{assumption}\label{parallel_trends}{(Sequential conditional parallel trends)}
\begin{align*}
    &\mathbb{E}\{Y_k(\bar{a}_{m-1},\underline{0})-Y_{k-1}(\bar{a}_{m-1},\underline{0})|\bar{L}_m,\bar{A}_{m-1}=\bar{a}_{m-1},A_m=1\}\\&=\mathbb{E}\{Y_k(\bar{a}_{m-1},\underline{0})-Y_{k-1}(\bar{a}_{m-1},\underline{0})|\bar{L}_m,\bar{A}_{m-1}=\bar{a}_{m-1},A_m=0\}.
\end{align*}
with probability one.
\end{assumption}
In words, this
assumption states that, conditional on history of adjustment variables through
time $m$ and treatment history through time $m-1$, the expected
counterfactual outcome trends under no treatment from time $m$ onwards
do not depend on the treatment actually received at time $m$. 

\begin{remark}\label{remark_pt}
    Here, we provide a condensed summary of Section 8.2 of \citet{us2022} discussing Assumption \ref{parallel_trends} and comparing it to the `never-treated' parallel trends assumption of \citet{callaway2021difference} in the staggered adoption setting. See the original source for a fuller discussion. In our notation, `never-treated' parallel trends states 
    \begin{align}
\begin{split}\label{pt_cands}
     &\mathbb{E}\{Y_k(\bar{0})-Y_{k-1}(\bar{0})|\bar{L}_m,\bar{A}_{m-1}=\bar{0}_{m-1},A_m=1\}\\&=\mathbb{E}\{Y_k(\bar{0})-Y_{k-1}(\bar{0})|\bar{L}_m,\bar{A}_K=\bar{0}\}.
     \end{split}
\end{align}
That is, given covariate history through time $m$, the expected counterfactual untreated future outcome trends in units that started treatment at time $m$ are equal to the expected outcome trends in units that in actuality never initiate treatment for the duration of follow up. When $\bar L$ contains a common cause $\bar Z$ of treatment and trends. The problem is that the distribution of $\underline{Z}_{m+1}(\bar{0})$ in the never treated units ($\bar A_K=0$) defining the conditioning event on the right hand side of (\ref{pt_cands}) would differ from the distribution of $\underline{Z}_{m+1}(\bar{0})$ in the units defining the conditioning event of the left hand side of (\ref{pt_cands}). This is because future values of the covariate $\underline{Z}_{m+1}(\bar{0})$ are associated with future treatment $\underline{A}_{m+1}$ conditioned on only in the never treated control group. As $\bar{Z}$ is also associated with the trends, the trends will therefore also differ across the two groups, violating the assumption (\ref{pt_cands}).

A sketch of a formal argument comparing (\ref{parallel_trends}) and (\ref{pt_cands}) is as follows. First, whenever Assumption \ref{parallel_trends} holds, (\ref{pt_cands}) will fail. This is because the left hand sides of the two assumptions are identical, but their right hand sides will differ due to conditioning on future treatments in (\ref{pt_cands}). Furthermore, if Assumption \ref{parallel_trends} fails, then (\ref{pt_cands}) will only hold if two loosely related quantities (namely, the magnitude of the violation of Assumption \ref{parallel_trends} and the magnitude of an association between future treatments and counterfactual outcome trends) are exactly equal. Thus, in the presence of a time-varying confounder $\bar{Z}$, (\ref{pt_cands}) can only hold by remarkable coincidence. When all confounders are pre-baseline, however, the two assumptions are actually essentially equivalent. The above discussion also applies to other parallel trends assumption variants, such as the `not yet treated' by the time of the outcome assumption \citep{callaway2021difference}, that also condition on future treatment values.

\citet{us2022} provide a Single World Intervention Graph (SWIG) \citep{richardson2013single} illustrating conditions under which Assumption \ref{parallel_trends} would hold in the presence of a time-varying confounder $\bar{Z}$, which we reproduce in Figure \ref{fig:did_swig}. In the SWIG, $\bar L_m= (\bar Z_m,\bar\Delta_m)$ and the future counterfactual untreated trends are sequentially exchangeable with treatment, which can be read off the graph by the rules of d-separation. The graph allows for an unmeasured common cause $U_{A}$ of treatment history $\bar A_K$ and the baseline outcome $Y_0$ as well as an unmeasured common cause $U_{\bar L}$ of baseline outcome $Y_0$ and $\bar L_K$. No unmeasured common causes of $\bar A_K$ and $\bar L_K$ are permitted, which implies that $U_A$ and $U_{\bar L}$ are marginally independent. Furthermore, $Y_0$ does not cause $\bar L_K$ or $\bar A_K$. Note that $Y_0$ could not be included in the conditioning set $\bar L$, as doing so would open the collider path $A_1\leftarrow U_{A}\rightarrow Y_0 \leftarrow U_{L}\rightarrow\Delta_2(0,0)$, thereby breaking identification. For comparison, we also provide a SWIG in Figure \ref{fig:did_swig_alt} lacking time-varying confounding under which both \eqref{pt_cands} and (\ref{parallel_trends}) would hold. 
\end{remark}

\begin{figure}
    \centering
\begin{tikzpicture}
\tikzset{line width=1.5pt, outer sep=0pt,
ell/.style={draw,fill=white, inner sep=2pt,line width=1.5pt},swig vsplit={gap=5pt,inner line width right=0.5pt}};
\node[name=uly, ell, shape=ellipse]{$U_{L}$};
\node[name=uay, right=5mm of uly, ell,  shape=ellipse]{$U_{A}$};
\node[name=a0,right=5mm of uay, shape=swig vsplit]{
\nodepart{left}{$A_0$}
\nodepart{right}{$0$} };
\node[name=l0, ell, shape=ellipse, above=10mm of a0,xshift=-10mm]{$Z_0$};
\node[name=y0, ell,  below=10mm of a0, ell,  shape=ellipse]{$Y_0$};
\node[name=y1, right=5mm of a0, ell, shape=ellipse]{$\Delta_1(0)$};
\node[name=a1,right=5mm of y1, shape=swig vsplit]{
\nodepart{left}{$A_1(0)$}
\nodepart{right}{$0$} };
\node[name=l1,  ell, shape=ellipse, above=10mm of y1, xshift=-10mm]{$Z_1(0)$};
\node[name=y2, right=5mm of a1, ell, shape=ellipse]{$\Delta_2(0,0)$};
\draw[->,line width=1.5pt,>=stealth,draw=gray!50]
(uly) edge[out=60,in=150] (l1)
(uly) edge (l0) 
(uly) edge[out=300,in=175] (y0)
(uly) edge[out=330,in=210] (y1)
(uly) edge[out=300,in=240] (y2)
(uay) edge (a0)
(uay) edge (y0)
(uay) edge[out=30,in=165] (a1)
;
\draw[->,line width=1.5pt,>=stealth]
(a0) edge (l1)
(a0) edge (y1)
(a0) edge[out=330,in=195] (a1)
(a0) edge[out=330,in=210] (y2)
(l0) edge (l1)
(l0) edge[out=260,in=150] (y0)
(l0) edge (y1)
(l0) edge (a1)
(l0) edge[out=400,in=120] (y2)
(l0) edge (a0)

(y1) edge[out=330,in=205] (y2)
(l1) edge (a1)
(l1) edge (y2)
(a1) edge (y2)
(l1) edge (y1)
;
\end{tikzpicture}
 \caption{Single World Intervention Graph (SWIG) depicting a scenario with general confounding structure under which our parallel trends assumption holds. Here, the adjustment set $\bar L_m=(\bar Z_m,\bar\Delta_m)$ with $\Delta_j$ denoting $Y_j-Y_{j-1}$. In the graph, treatments are d-separated from future counterfactual outcome increments given observed covariate history. Key encoded assumptions include: no unobserved common causes of $\bar A$ and $L$; relatedly, unobserved common causes $U_L$ of components of $L$ and unobserved common causes $U_A$ of treatment and the baseline outcome $Y_0$ are marginally independent; and the baseline outcome $Y_0$ does not have any descendants.}
 \label{fig:did_swig}
\end{figure}
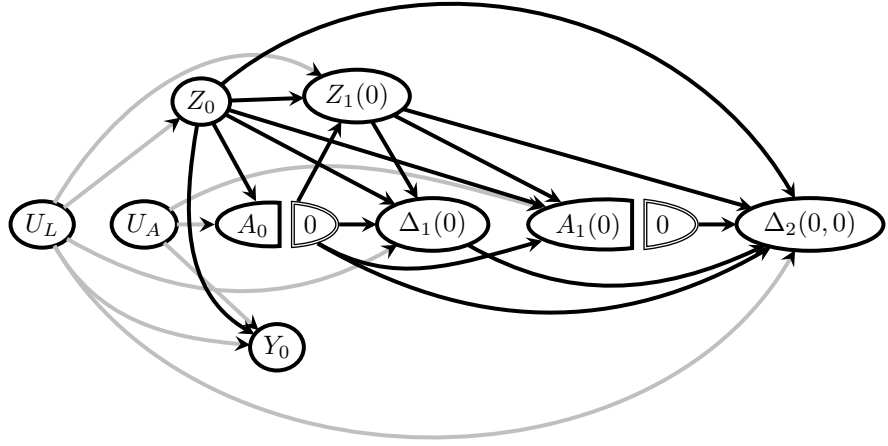

\begin{figure}
    \centering
\begin{tikzpicture}
\tikzset{line width=1.5pt, outer sep=0pt,
ell/.style={draw,fill=white, inner sep=2pt,line width=1.5pt},swig vsplit={gap=5pt,inner line width right=0.5pt}};
\node[name=uly, ell, shape=ellipse]{$U_{L}$};
\node[name=uay, right=5mm of uly, ell,  shape=ellipse]{$U_{A}$};
\node[name=a0,right=5mm of uay, shape=swig vsplit]{
\nodepart{left}{$A_0$}
\nodepart{right}{$0$} };
\node[name=l0, ell, shape=ellipse, above=10mm of a0,xshift=-10mm]{$Z_0$};
\node[name=y0, ell,  below=10mm of a0, ell,  shape=ellipse]{$Y_0$};
\node[name=y1, right=5mm of a0, ell, shape=ellipse]{$\Delta_1(0)$};
\node[name=a1,right=5mm of y1, shape=swig vsplit]{
\nodepart{left}{$A_1(0)$}
\nodepart{right}{$0$} };
\node[name=y2, right=5mm of a1, ell, shape=ellipse]{$\Delta_2(0,0)$};
\draw[->,line width=1.5pt,>=stealth,draw=gray!50]
(uly) edge (l0) 
(uly) edge[out=300,in=175] (y0)
(uly) edge[out=330,in=210] (y1)
(uly) edge[out=300,in=240] (y2)
(uay) edge (a0)
(uay) edge (y0)
(uay) edge[out=30,in=165] (a1)
;
\draw[->,line width=1.5pt,>=stealth]
(a0) edge (y1)
(a0) edge[out=330,in=195] (a1)
(a0) edge[out=330,in=210] (y2)
(l0) edge[out=260,in=150] (y0)
(l0) edge (y1)
(l0) edge (a1)
(l0) edge[out=400,in=120] (y2)
(l0) edge (a0)
(y1) edge[out=330,in=205] (y2)
(a1) edge (y2)
;
\end{tikzpicture}
 \caption{Single World Intervention Graph of a data generating process with only time-invariant confounding of trends in which assumptions \ref{parallel_trends} and \eqref{pt_cands} both hold. Here, $\bar L_m = Z_0$ for each $m$. As in Figure \ref{fig:did_swig}, $Y_0$ has no descendents and $U_L$ and $U_A$ are marginally independent.}
 \label{fig:did_swig_alt}
\end{figure}
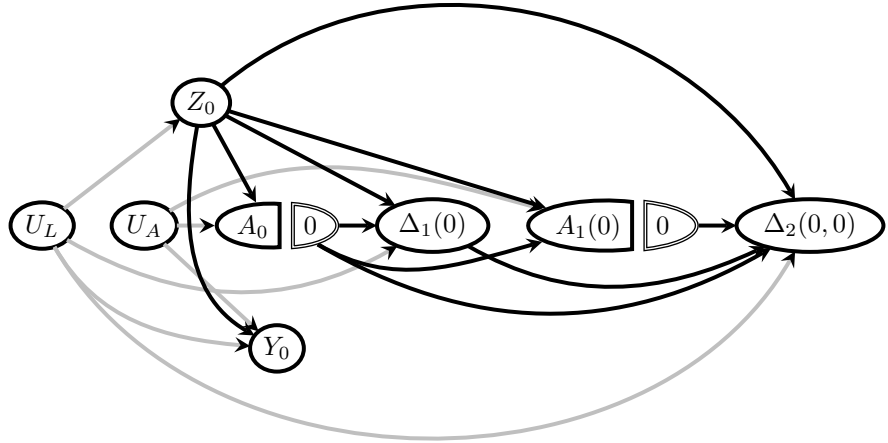

We further make sequential versions of the consistency assumption that counterfactual outcomes corresponding to observed treatments are equal to observed outcomes, and the positivity assumption. 
\begin{assumption}\label{consistency}{(Sequential consistency)}
\begin{align*}
Y_{m+1}(\bar{a}_{m}) = Y_{m+1} \text{ when }\bar{A}_m=\bar{a}_m.
\end{align*}
\end{assumption}
\begin{assumption}\label{positivity}{(Sequential positivity)}
\begin{align*}
    f_{A_m|\bar{L}_m,\bar{A}_{m-1}}(0|\bar{l}_m,\bar{a}_{m-1})>0 \text{ whenever } f_{\bar{L}_m,\bar{A}_{m-1}}(\bar{l}_m,\bar{a}_{m-1})>0.
\end{align*}
\end{assumption}
The identification derivation also makes use of the result that for increasing $\tilde{\bar{L}}_m$, without imposing any assumptions the consistency Assumption \ref{consistency}, blip functions satisfy the following important property \citep{jamie1994}:
\begin{equation}\label{blip_1994}
\mathbb{E}\left\{Y_k-\sum_{j=m}^{k-1}A_j\gamma_{jk}(\tilde{\bar{L}}_j,\bar{A}_{j-1})|\tilde{\bar{L}}_m,\bar{A}_m\right\}=\mathbb{E}\{Y_k(\bar{A}_{m-1},\underline{0})|\tilde{\bar{L}}_m,\bar{A}_m\}.
\end{equation}
For didactic purposes, in what follows we will derive nonparametric identification when $K=2$. In Theorem \ref{theorem_id_generalK} in the Appendix, we extend the result to general $K$ via induction.

When $K=2$, we have data $(L_0,Y_0,A_0,L_1,Y_1,A_1,Y_2)$ where $L_1$ may include $Y_0$. We can enumerate the blip functions (\ref{blip_models}) as follows:
\begin{align}
\gamma_{12}(\tilde{\bar{l}}_1,a_0)\equiv& \mathbb{E}\{Y_2(a_0,1)-Y_2(a_0,0)|\tilde{\bar{L}}_1=\tilde{\bar{l}}_1,A_0=a_0,A_1=1\}
\label{blips_2tp1}\\
\gamma_{01}(\tilde{l}_0)\equiv& \mathbb{E}\{Y_1(1)-Y_1(0)|\tilde{L}_0=\tilde{l}_0,A_0=1\}\label{blips_2tp2}\\
\gamma_{02}(\tilde{l}_0)\equiv& \mathbb{E}\{Y_2(1,0)-Y_2(0,0)|\tilde{L}_0=\tilde{l}_0,A_0=1\}.\label{blips_2tp3}
\end{align}
We can also enumerate the parallel trends assumptions: 
\begin{align}
\begin{split}\label{pt_2tp}
    \mathbb{E}\{Y_2(a_0,0)-Y_{1}|\bar{L}_1,A_0=a_0,A_1=1\}=&\mathbb{E}\{Y_2(a_0,0)-Y_{1}|\bar{L}_1,A_0=a_0,A_1=0\}\\
    \mathbb{E}\{Y_2(0,0)-Y_{1}(0)|L_0,A_0=1\}=&\mathbb{E}\{Y_2(0,0)-Y_{1}(0)|L_0,A_0=0\}\\
    \mathbb{E}\{Y_1(0)-Y_0|L_0,A_0=1\}=&\mathbb{E}\{Y_1(0)-Y_0|L_0,A_0=0\}.
    \end{split}
\end{align}
which all hold with probability one. Finally, the key instance of the property (\ref{blip_1994}) is that
\begin{align}
\mathbb{E}\{Y_2-A_1\gamma_{12}(\tilde{\bar{L}}_1,A_0)|\tilde{\bar{L}}_1,\bar{A}_1\}=\mathbb{E}\{Y_2(A_0,0)|\tilde{\bar{L}}_1,\bar{A}_1\}.
\end{align}
\begin{theorem}\label{id_2tp} \textbf{Two Time-Point Identification}. Under Assumptions \ref{parallel_trends}-\ref{positivity}, the blip functions (\ref{blips_2tp1}-\ref{blips_2tp3}) are identified as:
\begin{align*}
\gamma_{01}(\tilde{l}_0)=&\mathbb{E}(Y_1-Y_0|\tilde{L}_0 =\tilde{l}_0,A_0=1) - \mathbb{E}\{\mathbb{E}(Y_1-Y_0|L_0,A_0=0)|\tilde{L}_0=\tilde{l}_0,A_0=1\};\\
\gamma_{12}(\tilde{\bar{l}}_1,a_0)=&\mathbb{E}(Y_2-Y_1|\tilde{\bar{L}}_1=\tilde{\bar{l}}_1,A_0=a_0,A_1=1)\\& - 
    \mathbb{E}\{\mathbb{E}(Y_2-Y_1|\bar{L}_1 =\bar{l}_1,A_0=a_0,A_1=0)|\tilde{\bar{L}}_1=\tilde{\bar{l}}_1,A_0=a_0,A_1=1\};\\\gamma_{02}(\tilde{l}_{0})=&\mathbb{E}[Y_2-A_1\gamma_{12}(\tilde{\bar{L}}_{1},A_0)-\{Y_1-A_0\gamma_{01}(\tilde{l}_0)\}|\tilde{L}_0=\tilde{l}_0,A_0=1] \\
    &-\mathbb{E}[\mathbb{E}\{\mathbb{E}(Y_2-Y_1|\bar{L}_1,A_0=0,A_1=0)|L_0,A_0=0\}|\tilde{L}_0=\tilde{l}_0,A_0=1]
\end{align*}
\end{theorem}
\begin{proof}
See Appendix A.
\end{proof}

In Appendix A, we also give a proof by induction of identification of blip functions (\ref{blip_models}) for general number of time points.  Importantly, (\ref{blip_1994}) implies that when the blip functions (\ref{blip_models}) are identified, so too is  $\mathbb{E}\{\bar{Y}_K(\bar{0})\}$, the expected outcome trajectory under no treatment. Specifically, $\mathbb{E}\{Y_k(\underbar{0})\}=\mathbb{E}\{Y_k-\sum_{j=0}^{k-1}A_j\gamma_{jk}(\tilde{\bar{L}}_j,\bar{A}_{j-1})\}$ for $k$ in $(1,\ldots,K)$. Thus, identification of the blip functions (\ref{blip_models}) enables identification of the total population level impact of observed interventions compared to if no interventions had occurred, i.e. $\mathbb{E}\{\bar{Y}_K-\bar{Y}_K(\bar{0})\}$. 


\subsection{Staggered Adoption: an Important Special Case}\label{section_staggered}
Most time-varying DiD methods have been developed for staggered adoption settings where all units start off untreated and then begin treatment at different times. In many applications, initial treatment is an absorbing state; once direct treatment is initiated, it is maintained throughout follow up. An example of an absorbing binary treatment is Medicaid expansion, which we analyze in our real data application in Section \ref{application}. DiD methods for staggered adoption then generally target the estimand 
\begin{align}\label{equation:staggered_1}
\mathbb{E}\{Y_k(\bar{a}_{m-1}=\bar{0},\underline{1})-Y_k(\bar{0})|\bar{A}_{m}=(\bar{0}_{m-1},1)\};
\end{align}
i.e. the effect of initiating and continuing treatment versus remaining off treatment \citep{callaway2021difference}. However, in certain cases, units may switch on and off of treatment over the course of follow up. In that case, existing approaches then target the effect of initiation versus remaining off treatment, marginalizing over future treatment patterns under usual care:
\begin{align}\label{equation:staggered_2}
\mathbb{E}\{Y_k(\bar{a}_{m-1}=\bar{0},a_m=1)-Y_k(\bar{0})|\bar{A}_{m}=(\bar{0}_{m-1},1)\}.    
\end{align}

The methodology described in this paper can be straightforwardly adapted for either context. If one codes exposure such that it takes the value 1 only at the time of first treatment and 0 at all other times, then, in the special case that $\tilde{\bar{L}}_m$ is the empty set, the blip functions (\ref{blip_models}) represent the effects of treatment initiation, but not necessarily continuation, in the treated. Specifically, under this `first treatment' coding, $\gamma_{mk}(\bar{a}_{m-1}=\bar{0})$ equals \eqref{equation:staggered_1} or \eqref{equation:staggered_2} (depending on whether treatment is absorbing). Of course, we do not restrict $\tilde{\bar{L}}_m$ to be the empty set and thus enable investigation of staggered adoption effects conditional on evolving covariate history. In our Medicaid application in Section \ref{application}, by estimating (\ref{blip_models}) with `first treatment' coding, we gain insight into how the effects of Medicaid expansion on uninsurance rates depend on unemployment rates prior to expansion. 

When treatment is non-absorbing, there may be still interested in inferring the effect of the sustained strategy as in \eqref{equation:staggered_1}; we discuss how this can be done using mSNMMs in Section \ref{section_general}

\subsection{Semiparametric Estimation}

As with our identification results, we will present our semiparametric theory results for two time points in the main text with generalizations to arbitrary number of time points given in Appendix B. We will assume for each possible value $\tilde{l}_0$ of $\tilde{L}_0$, there exist known functions $\gamma_{01}(\tilde{l}_0;\cdot):\Psi \mapsto \mathbb{R}$ and $\gamma_{02}(\tilde{l}_0;\cdot):\Psi \mapsto \mathbb{R}$, $0\in \Psi \subset \mathbb{R}^p$; they both satisfy $\gamma_{01}(\tilde{l}_0;0)=0$ and $\gamma_{02}(\tilde{l}_0;0)=0$, such that $\gamma_{01}(\tilde{l}_0)=\gamma_{01}(\tilde{l}_0;\psi^*)$ and
$\gamma_{02}(\tilde{l}_0)=\gamma_{02}(\tilde{l}_0;\psi^*),$ where $\psi^*\in \Psi$ is unknown. Likewise, for each possible value $\tilde{\bar{l}}_1$ of $\tilde{\bar{L}}_1$ and $a_0$ of $A_0$, there exists a known function $\gamma_{12}(\tilde{\bar{l}}_1,a_0;\cdot):\Psi \mapsto \mathbb{R}$, which satisfies $\gamma_{12}(\tilde{\bar{l}}_1,a_0;0)=0$, such that $\gamma_{12}(\tilde{\bar{l}}_1,a_0)=\gamma_{02}(\tilde{\bar{l}}_1,a_0;\psi^*)$. Adopting the reasoning from the previous section, we define the model $\mathcal{M}$ to denote the collection of distributions $P$ for which 
\begin{align*}
\gamma_{01}(\tilde{l}_0;\psi)=&\mathbb{E}_P(Y_1-Y_0|\tilde{L}_0 =\tilde{l}_0,A_0=1) - \mathbb{E}_P\{\mathbb{E}_P(Y_1-Y_0|L_0,A_0=0)|\tilde{L}_0=\tilde{l}_0,A_0=1\};\\
\gamma_{12}(\tilde{\bar{l}}_1,a_0;\psi)=&\mathbb{E}_P(Y_2-Y_1|\tilde{\bar{L}}_1=\tilde{\bar{l}}_1,A_0=a_0,A_1=1)\\& - 
    \mathbb{E}_P\{\mathbb{E}_P(Y_2-Y_1|\bar{L}_1 =\bar{l}_1,A_0=a_0,A_1=0)|\tilde{\bar{L}}_1=\tilde{\bar{l}}_1,A_0=a_0,A_1=1\};\\\gamma_{02}(\tilde{l}_{0};\psi)=&\mathbb{E}_P[Y_2-A_1\gamma_{12}(\tilde{\bar{L}}_{1},A_0;\psi)-\{Y_1-A_0\gamma_{01}(\tilde{l}_0;\psi)\}|\tilde{L}_0=\tilde{l}_0,A_0=1] \\
    &-\mathbb{E}_P[\mathbb{E}_P\{\mathbb{E}_P(Y_2-Y_1|\bar{L}_1,A_0=0,A_1=0)|L_0,A_0=0\}|\tilde{L}_0=\tilde{l}_0,A_0=1]
\end{align*}
for some $\psi \in \Psi$. We assume $\gamma_{01}(\cdot;\cdot), \gamma_{01}(\cdot;\cdot), \gamma_{12}(\cdot;\cdot), \gamma_{02}(\cdot;\cdot)$ are all smooth in $\psi$. Let $\pi^*_1(\bar{L}_1,A_0)\equiv \mathbb{E}(A_1|\bar{L}_1,A_0)$, $\pi_0^*(L_0)\equiv\mathbb{E}(A_0|L_0)$, $\mu^*_{01}\equiv \mathbb{E}(Y_1-Y_0|L_0,A_0=0)$, $\mu^*_{12}(\bar{L}_1,A_0) \equiv \mathbb{E}(Y_2-Y_1|\bar{L}_1,A_0,A_1=0)$ and $\mu^*_{02}(L_0) \equiv \mathbb{E}\{\mu^*_{12}(\bar{L}_1,A_0)|L_0,A_0=0\}$. We will sometimes use the shorthand $\pi^*\equiv (\pi^*_0,\pi^*_1)$ and $\mu^*\equiv (\mu^*_{01},\mu^*_{02},\mu^*_{12})$, and likewise for the estimators.
\begin{theorem}\label{IF_2times}
The set of influence functions for all RAL estimators of $\psi^*$ under the model $\mathcal{M}$ is given by
\[\left\{D(\psi^*,q_0,q_1)^{-1}\{\phi_0(O;\psi^*,q_0,\pi^*,\mu^*)+\phi_1(O;\psi^*,q_1,\pi^*,\mu^*)\}:q_0,q_1\textrm{ unrestricted}\right\}\cap \mathcal{L}_2
\]
where$q_{0}(\tilde{L}_0)$ and $q_{1}(\tilde{\bar{L}}_1,A_0)$ are, respectively, $p \times 2$-dimensional and $p$-dimensional arbitrary functions, such that $D(\psi^*,q_0,q_1)^{-1}$ exists. Also, 
 for any $\psi\in\Psi$, and any arbitrary functions $\pi_0(L_0)$, $\pi_1(\bar{L}_1,A_0)$, $\mu_0(L_0)$ and $\mu_1(\bar{L}_1,A_0)$ (where $\pi_0(L_0), \pi_1(\bar{L}_1,A_0)\neq1$ with probability one),
\begin{align*}
\phi_0(O;\psi,q_0,\pi,\mu) \equiv & q_{0}(\tilde{L}_0)\epsilon_0(O;\psi,\pi,\mu), \\
\phi_1(O;\psi,q_1,\pi,\mu) \equiv & q_{1}(\tilde{\bar{L}}_1,A_0)\epsilon_1(O;\psi,\pi,\mu), \\
D(\psi,q_0,q_1)\equiv& -\frac{\partial }{\partial \psi}\mathbb{E}\left\{q_{0}(\tilde{L}_0)\epsilon_0(O;\psi,\pi,\mu)+q_{1}(\tilde{\bar{L}}_1,A_0)\epsilon_{1}(O;\psi,\pi,\mu)\right\}|_{\psi=\psi}
\end{align*}
with
$\epsilon_{0}(O;\psi,\pi,\mu)=(\epsilon_{01}(O;\psi,\pi,\mu),\epsilon_{02}(O;\psi,\pi,\mu))^T$ and
\begin{align*}
\epsilon_{01}(O;\psi,\pi,\mu)\equiv&\frac{\{A_0-\pi_0(L_0)\}}{\{1-\pi_0(L_0)\}}\{Y_1-Y_0-\mu_{01}(L_0)\}-A_0\gamma_{01}(\tilde{L}_0;\psi) ,\\
\epsilon_{02}(O;\psi,\pi,\mu)\equiv&\left[1-\frac{(1-A_1)(1-A_0)}{\{1-\pi_1(\bar{L}_1,A_0)\}\{1-\pi_0(L_0)\}}\right](Y_2-Y_1)\\
    &-\frac{(1-A_0)}{\{1-\pi_0(L_0)\}}\left[1-\frac{1-A_1}{\{1-\pi_1(\bar{L}_1,A_0)\}}\right]\mu_{12}(\bar{L}_1,A_0)-\left[1-\frac{(1-A_0)}{\{1-\pi_0(L_0)\}}\right]\mu_{02}(L_0)\\
    &-A_1\gamma_{12}(\tilde{\bar{L}}_1,A_0;\psi)+A_0\gamma_{01}(\tilde{L}_0;\psi)-A_0\gamma_{02}(\tilde{L}_0;\psi),\\
    \epsilon_{1}(O;\psi,\pi,\mu)\equiv& \frac{\{A_1-\pi_1(\bar{L}_1,A_0)\}}{\{1-\pi_1(\bar{L}_1,A_0)\}}\{Y_2-Y_1-\mu_{12}(\bar{L}_1,A_0)\}-A_1\gamma_{12}(\tilde{\bar{L}}_1,A_0;\psi).
\end{align*}
\end{theorem}
The proof follows from a more general result (for arbitrary numbers of time points) given in the Appendix.

Theorem \ref{IF_2times} can be used to construct cross-fit, DR-ML estimators of $\psi^*$ similar to Section 2.  In Algorithm \ref{algorithm:cfdrml-2t}, we describe how to do this with $S$-fold splitting. The following theorem outlines the asymptotic properties of the cross-fit estimator. 

\begin{algorithm}
\caption{Implementation of a cross-fit DR-ML estimator: time-varying setting}\label{algorithm:cfdrml-2t}
 Randomly split the $N$ units into $S$ disjoint folds of size $n_s\approx N/S$.\\
For each $s=1,...,S$, construct estimators $\hat{\mu}^{(-s)}$ of $\mu^*$ and $\hat{\pi}^{(-s)}$ of $\pi^*$, using all the units except those in fold $s$.\\
For each $s=1,...,S$, compute $\hat{\psi}^{(s)}$ from the $s$-th fold as the (assumed unique) solution to the estimating equations $\mathbb{P}_{n_s}\{\phi_0(O;\psi,q_0,\hat{\pi}^{(-s)},\hat{\mu}^{(-s)})+\phi_1(O;\psi,q_1,\hat{\pi}^{(-s)},\hat{\mu}^{(-s)})\}=0$ where $\mathbb{P}_{n_s}$ refers here to the empirical measure in the $s$-th fold.\\
Take the weighted average of the $S$ estimators:  $\hat{\psi}^{cf}=\sum^{S}_{s=1}\left(\frac{n_s}{N}\right)\hat{\psi}^{(s)}$.
\end{algorithm}

\begin{theorem}\label{theorem:estimation_2t}
Suppose the following conditions hold:
\begin{enumerate}
    \item The map $\psi \mapsto \mathbb{P}\{\phi_0(o;\psi,q_0,\pi,\mu)+\phi_1(o;\psi,q_1,\pi,\mu)\}$ is differentiable at $\psi^*$ uniformly in $(\pi,\mu)$. 
    \item The matrix $D(\psi^*,q_0,q_1)$ is non-singular.
    \item $\{1-\hat{\pi}^{(-s)}_0(L_0)\}\geq c_0>0$, $\{1-\pi^*_0(L_0)\}\geq c_0>0$, $\{1-\hat{\pi}^{(-s)}_1(\bar{L}_1,A_0)\}\geq c_1>0$ and $\{1-\pi^*_1(\bar{L}_1,A_0)\}\geq c_1>0$ with probability one for constants $c_0,c_1$ and $s=1,...,S$.
    \item For a chosen $p\times 2$ function $q_0(\tilde{L}_0)$, the maximum absolute value of any element in $q_0(\tilde{L}_0)$ is upper bounded by $C_0<\infty$ with probability one. Likewise, for a chosen  $p\times 1$ function $q_1(\tilde{\bar{L}}_1)$, the maximum absolute value of any element in $q_1(\tilde{\bar{L}}_1)$ is upper bounded by $C_1<\infty$ with probability one.
    \item $\hat{\psi}^{(s)}-\psi^*=o_{\mathbb{P}}(1)$, $\norm{\hat{\pi}^{(-s)}_{0} - \pi_{0}^*}=o_{\mathbb{P}}(1)$, $\norm{\hat{\mu}^{(-s)}_{01}-\mu_{01}^*}=o_{\mathbb{P}}(1)$, and $\norm{\hat{\mu}^{(-s)}_{02}-\mu_{02}^*}=o_{\mathbb{P}}(1)$ and $\norm{\hat{\mu}^{(-s)}_{12}-\mu_{12}^*}$ for $s=1,...,S$.
    \item 
    \begin{align*}
    \norm{\hat{\pi}^{(-s)}_{0} - \pi_{0}^*}\norm{\hat{\mu}^{(-s)}_{01}-\mu_{01}^*}=& o_\mathbb{P}(n^{-1/2})\\
      \norm{\hat{\pi}^{(-s)}_{0} - \pi_{0}^*}\norm{\hat{\mu}^{(-s)}_{02}-\mu_{02}^*}=& o_\mathbb{P}(n^{-1/2})\\
       \norm{\hat{\pi}^{(-s)}_{1} - \pi_{1}^*}\norm{\hat{\mu}^{(-s)}_{12}-\mu_{12}^*}=& o_\mathbb{P}(n^{-1/2})
    \end{align*} for $s=1,...,S$.
    \item For any fixed ($\pi,\mu$), the class $\left\{\phi_0(o;\psi,q_0,\pi,\mu)+\phi_1(o;\psi,q_1,\pi,\mu):\psi\in \Psi\right\}$ is Donsker in $\psi$. 
\end{enumerate}
Then $\hat{\psi}^{cf}$ satisfies the expansion
\begin{equation*}
\sqrt{N}(\hat{\psi}^{cf}-\psi^{*})=D(\psi^*,q_0,q_1)^{-1} \frac{1}{\sqrt{N}}\sum_{i=1}^N\{\phi_0(O;\psi^*,q_0,\pi^*,\mu^*)+\phi_1(O;\psi^*,q_1,\pi^*,\mu^*)\} +o_\mathbb{P}(1).
\end{equation*}
\end{theorem}

We close this section with a result regarding the efficiency of our estimators.
\begin{theorem}\label{theorem:eff_score_2t}
The efficient influence function for $\psi^*$ under model $\mathcal{M}$ is given by
\begin{align*}
D(\psi^*,q_{eff,0},q_{eff,1})^{-1}[q_{eff,0}(\tilde{L}_0)\epsilon_0(O;\psi^*,\pi^*,\mu^*)+q_{eff,1}(\tilde{\bar{L}}_1,A_0)\epsilon_1(O;\psi^*,\pi^*,\mu^*)]
\end{align*}
where 
\begin{align*}
q_{eff,0}(\tilde{L}_0)=&-\mathbb{E}\left\{\frac{\partial}{\partial\psi}\mathfrak{e}_0(O;\psi,\pi^*,\mu^*)|_{\psi=\psi^*}|\tilde{L}_0\right\}^T\mathbb{E}\{\mathfrak{e}_0(O;\psi^*,\pi^*,\mu^*)\mathfrak{e}_0(O;\psi^*,\pi^*,\mu^*)^T|\tilde{L}_0\}^{-1}\\
q_{eff,1}(\tilde{\bar{L}}_1,A_0)=&-\mathbb{E}\left\{\frac{\partial}{\partial\psi}\epsilon_1(O;\psi,\pi^*,\mu^*)|_{\psi=\psi^*}|\tilde{\bar{L}}_1,A_0\right\}^T\mathbb{E}\{\epsilon_1(O;\psi^*,\pi^*,\mu^*)^2|\tilde{\bar{L}}_1,A_0\}^{-1}\\
&-q_{eff,0}(\tilde{L}_0)\mathbb{E}\{\epsilon_0(O;\psi^*,\pi^*,\mu^*)\epsilon_1(O;\psi^*,\pi^*,\mu^*)|\tilde{\bar{L}}_1,A_0\}\mathbb{E}\{\epsilon_1(O;\psi^*,\pi^*,\mu^*)^2|\tilde{\bar{L}}_1,A_0\}^{-1}\\
\mathfrak{e}_0(O;\psi^*,\pi^*,\mu^*)=&\epsilon_0(O;\psi^*,\pi^*,\mu^*)\\&-\mathbb{E}\{\epsilon_0(O;\psi^*,\pi^*,\mu^*)\epsilon_1(O;\psi^*,\pi^*,\mu^*)|\tilde{\bar{L}}_1,A_0\}\mathbb{E}\{\epsilon_1(O;\psi^*,\pi^*,\mu^*)^2|\tilde{\bar{L}}_1,A_0\}^{-1}\epsilon_1(O;\psi^*,\pi^*,\mu^*).
\end{align*}
\end{theorem}
Theorem \ref{theorem:eff_score_2t} is currently more of theoretical interest rather than practical, as the efficient score may be challenging to operationalise in practice. This is especially the case as the number of measurement times grows large, due to the complex recursive expressions involved.

\section{General Regime SNMMs}\label{section_general}
There is not necessarily anything special about the untreated regime. We might also be interested in the conditional effects of a blip of treatment followed by adherence to an arbitrary regime $g$ compared to following $g$ at the current time and thereafter \citep{jamie2004}. $g$ could be a dynamic regime, i.e. $%
g=(g_0,g_1,\ldots,g_K)$ a vector of functions $g_t: (\tilde{\bar{L}}_t,\bar{%
A}_{t-1}) \rightarrow a_t$ that determine treatment values at each time
point $t$ based on $(\bar{A}_{t-1},\tilde{\bar{L}}_t)$. Blip functions representing causal contrasts of interest would be
\begin{equation}\label{g_blip}
    \gamma_{mk}^g(\tilde{\bar{l}}_m,\bar{a}_{m})\equiv \mathbb{E}\{Y_k(\bar{a}_{m},\underline{g}_{m+1})-Y_k(\bar{a}_{m-1},\underline{g}_m)|\tilde{\bar{L}}_m=\tilde{\bar{l}}_m,\bar{A}_{m-1}=\bar{a}_{m}\}.
\end{equation}
These represent the effects of a blip of treatment at time $m$ at level $a_m$ followed by regime $g$ thereafter compared to regime $g$ at time $m$ and thereafter, conditional on treatment and covariate history through time $m$. In this setting, we make the parallel trends assumption
\begin{align}
\begin{split}\label{parallel_g}
    &\mathbb{E}\{Y_k(\bar{a}_{m-1},\underline{g}_m)-Y_{k-1}(\bar{a}_{m-1},\underline{g}_m)|\bar{L}_m, \bar{A}_m=\bar{a}_{m}\}\\
   &=\mathbb{E}\{Y_k(\bar{a}_{m-1},\underline{g}_m)-Y_{k-1}(\bar{a}_{m-1},\underline{g}_m)|\bar{L}_m,\bar{A}_m=(\bar{a}_{m-1},g(\bar{a}_{m-1},\tilde{\bar{L}}_m))\}.
\end{split}
\end{align}
That is, we assume that the counterfactual future outcome trends under $g$ are conditionally mean independent of current treatment given covariate and treatment history. The identification and estimation results obtained earlier for $g=0$ go through simply substituting in a general $g$, though the meaning of the parallel trends assumption varies with $g$. Further, as in the case of $g=0$, with identification of $\gamma_{mk}^g(\tilde{\bar{l}}_m,\bar{a}_{m-1})$ comes identification of the expected counterfactual outcome trajectory $\mathbb{E}\{\bar{Y}_K(g)\}$. 

One particular dynamic regime of interest might be the `continue previous treatment' regime
\begin{equation}\label{continue_g}
g^{cont}(\bar{A}_{m-1},\tilde{\bar{L}}_m)=A_{m-1}.
\end{equation}
The blip function corresponding to this regime is  
\begin{equation}\label{cont_est}
\gamma_{mk}^{g^{cont}}(\tilde{\bar{l}}_m,\bar{a}_{m})= \mathbb{E}\{Y_k(\bar{a}_{m-1},\underline{1})-Y_k(\bar{a}_{m-1},\underline{0})|\tilde{\bar{L}}_m=\tilde{\bar{l}}_m,\bar{A}_{m}=\bar{a}_{m}\}.
\end{equation}
For $a_m=1$ and $a_{m-1}=0$, (\ref{cont_est}) is the (conditional) effect on the outcome at time $k>m$ of (re-)starting and continuing treatment from time $m$ versus never (re-)starting treatment in those who do (re-)start treatment at time $m$. An issue that has caused some consternation in the DiD literature is how to estimate effects of sustained interventions like (\ref{cont_est}) when units might switch on and off treatment in the data; see section 3.4 of \citet{roth2023}. Due to limitations in available methods, investigators have had to settle for the effect of initiation marginalizing over subsequent treatment patterns under usual care, as discussed in Section \ref{section_staggered}. Under the parallel trends assumption (\ref{parallel_g}) corresponding to $g^{cont}$, (\ref{cont_est}) is identified and can be estimated by solving the estimating equations of the previous section with $\underline{g}^{cont}(\bar{a}_{m-1})$ and $\underline{g}^{cont}(\bar{a}_{m})$ replacing $\underline{0}_m$ and $\underline{0}_{m+1}$, respectively. 

 \citet{renson2023identifying} enable estimation of the expected counterfactual outcome under a single sustained treatment trajectory $E[\bar{Y}_K(\bar{a})]$, e.g. $E[\bar{Y}_K(\bar{0}_{t-1},\underline{1}_t)]$ for a specific time $t$. However, even applying their approach to multiple time points $t$ does not yield estimates of the causal \textit{contrasts} (\ref{cont_est}) of interest. 


\section{Simulations}\label{section_sim}

We generated 1,000 data sets of 20,000 observations each from the following data generating process, which we modified from a SNMM simulation in \citet{almirall2012designing}:
\begin{align*}
    &\mathbb{E}(Y_0)=\mathbb{E}\{Y_1(0)\}=\mathbb{E}\{Y_2(0,0)\}=17;\\
    &U\sim N(0,1); Y_0\sim N(17+U,1); L_0^{(1)}\sim N(0.5,0.82); L_0^{(2)}\sim N(0.5,0.82); \\
    &A_0\sim Bernoulli(logit^{-1}(0.5-0.5L_0^{(1)}-0.5L_0^{(2)}+U));\\
    &e_{01}(L_0^{(1)},L_0^{(2)})=0.05(L_0^{(1)}+L_0^{(2)})-0.05; Y_1\sim N(17+e_{01}(L_0^{(1)},L_0^{(2)})+A_0\gamma_{01}(L_0^{(2)})+U,1);\\
    &L^{(1)}_1,L^{(2)}_1\sim N(
        0.5 + 0.05(L_0^{(1)}+L_0^{(2)}) - 0.5 A_0 + 0.35L_0^{(1)} A_0, 0.65)\\
    & A_1 =Bernoulli(logit^{-1}(0.5 - 0.5 (L_1^{(1)}+L_1^{(2)}) + U));\\
    &e_{02}(L_0^{(1)},L_0^{(2)})=0.25(L_0^{(1)}+L_0^{(2)}-0.5); e_{12}(\bar{L}^1_1,\bar{L}_1^{(2)},A_0)= 0.5(L_1^{(1)}+L_1^{(2)}) -0.25;\\
    &Y_2\sim N(17+e_{02}(L_0^{(1)},L_0^{(2)})+e_{12}(\bar{L}^{(1)}_1,\bar{L}_1^{(2)},A_0)+A_0\gamma_{02}(L_0^{(2)})+A_1\gamma_{12}(\bar{L}^{(2)}_1)+U,1)
\end{align*}
For data generated as above, the functions $\gamma_{mk}(\bar{L}^{(2)}_m)$ represent the blip function contrasts (\ref{blip_models}). (\citet{almirall2012designing} provide a justification of this point via a decomposition of the likelihood containing the blip functions.) We specified that 
\begin{align*}
    \gamma_{01}(L_0^{(2)})=.45+.45L_0^{(2)}\\
    \gamma_{12}(\bar{L}^{(2)}_1,A_0)=.45+.45L_1^{(2)}\\
    \gamma_{02}(L_0^{(2)})=.45+.45L_0^{(2)}
\end{align*}
 Because the unobserved confounder $U$ enters into the structural equations for the untreated outcomes in a constant additive fashion, it is straightforward to show that the parallel trends Assumption \ref{parallel_trends} holds at each time point conditional on the history of the adjustment set $(\bar{L}^{(1)},\bar{L}^{(2)})$. We model heterogeneity as a function of $\bar{L}^2$ alone, so that it plays the role of $\tilde{\bar{L}}$. We specified correct blip models 
\begin{equation*}
    \gamma_{mk}(\bar{L}_m^{(2)};\psi)=\psi^{(1)}_{mk}+\psi^{(2)}_{mk}.
\end{equation*}
We estimated $\psi^*$ in two ways. First, we constructed cross-fit estimators $\hat{\psi}^{cf}$ following Algorithm \ref{algorithm:cfdrml-2t} with two equal sized folds and using gradient boosted trees with the default tuning parameters of the R package xgboost \citep{chen2019package} for all nuisance regressions. We chose index functions $\{q_{01}=(1,L_0^{(2)}), q_{12}=(1,L_1^{(2)}),q_{02}=(1,L_0^{(2)})\}$. We also computed a parametric estimator $\hat{\psi}^{par}$ solving $\mathbb{P}_n\{\phi_0(O;\psi,q_0,\hat{\pi},\hat{\mu})+\phi_1(O;\psi,q_1,\hat{\pi},\hat{\mu})\}=0$ in the full sample with the same index functions as the first estimator, foregoing cross-fitting. For the parametric estimator, we used generalized linear models with 3-knot spline basis expansions of all continuous predictors for the nuisance regressions. Under our data generating process, the treatment nuisance models were correctly specified in this parametric estimator. For both estimators, we constructed 95\% confidence intervals based on plug-in estimates of the sandwich variance estimator. R code for the simulation can be found at https://github.com/zshahn/marginal-structural-nested-mean-models-under-parallel-trends.    
Both estimators were approximately unbiased and exhibited good coverage  (Table \ref{tab:tv_sim}). The parametric estimator, unsurprisingly, was slightly more efficient, but the cross-fit estimator was competitive despite not requiring a correct parametric nuisance model specification.


\begin{table}\label{tv_sim}
\centering
\begin{tabular}{ |p{2.4cm}|p{1.5cm}|p{1.5cm}|p{1.5cm}|p{1.5cm}|p{1.5cm}|p{1.5cm}|  }
 \hline
 & $\psi^{01}_1=.45$  &$\psi^{01}_2=.45$ &$\psi^{12}_1=.45$ &$\psi^{12}_2=.45$ &$\psi^{02}_1=.45$ &$\psi^{02}_2=.45$\\
 \hline
 Cross-fit DML   & 0.00 (0.94) [0.04] &  0.00 (0.95) [0.04]& 0.01 (0.95) [0.04]& -0.01 (0.96) [0.06] & -0.01 (0.92) [0.07] & 0.00 (0.94) [0.09]\\
 \hline
 Parametric& 0.00 (0.93) [0.03] &  0.00 (0.94) [0.03]& 0.00 (0.95) [0.03]& 0.00 (0.94) [0.04] & -0.00 (0.94) [0.04] & 0.00 (0.95) [0.05]\\
 \hline
 \end{tabular}
 \caption{Bias (coverage of 95\% CIs) [sampling standardard deviation] of parameter estimates over 1,000 simulated data sets of 20,000 observations each}
    \label{tab:tv_sim}
\end{table}

\section{Real Data Application: Effects of Medicaid Expansion on Uninsurance Rates}\label{application}
Medicaid is a government funded program in the United States offering health insurance coverage to uninsured individuals meeting income eligibility or other criteria. Starting in 2014, the Affordable Care Act gave states the opportunity to receive federal subsidies to expand access to nearly all adults with incomes under 138$\%$ of the federal poverty line. States have taken advantage of this opportunity and expanded coverage in a staggered fashion. Past studies have found heterogeneity in Medicaid expansion effects, e.g. that Medicaid expansion was more impactful in rural areas \citep{antonisse2018effects}.  

As an illustrative analysis, we use marginal SNMMs to study how the effects of expansions in 2014 and 2015 on county level uninsurance rates over the same period (among adults with incomes under 138$\%$ of the federal poverty line) varied with county unemployment rate in the previous year. Because no state has retracted an expansion, we are interested in effects of initiation and code initial treatment as 1 and subsequent treatments as 0 as discussed in Section \ref{section_staggered}. 

We specified the blip models
\begin{equation*}
    \gamma_{mk}(\bar{L}^{unemp}_m;\psi^*_{mk})=\mathbb{E}\{Y_k-Y_k(\bar{0})|A_m=1,\bar{L}^{unemp}_m\}=\psi^*_{mk0}+\psi^{*}_{mk1}L^{unemp}_m,
\end{equation*}
for $k\geq m$, $m\in\{2014,2015\}$, and $k\in \{2014, 2015, 2016\}$. To be consistent with the temporal ordering throughout the paper, $Y_k$ is the uninsurance rate over the previous year (e.g. $Y_{2014}$ is the uninsurance rate in the 2013-2014 period), $A_m$ indicates whether a county's state expanded Medicaid for the first time in year $m$, and $L_m^{unemp}$ indicates unemployment rate in year $m-1$. For simplicity, we treat all expansions as if they occurred at the start of the calendar year (e.g. $A_{2014}=1$ indicates that a state expanded at the start of 2014), which almost all did. These blip models are flexible in allowing the parameters to vary arbitrarily with each treatment and outcome year pair. However, they still make a number of simplifying assumptions. One could consider more flexible specifications (e.g. including covariate history from earlier years and spline or polynomial terms) as sample size allows.

We assume that parallel trends holds conditional on our effect modifier of interest (prior year unemployment) along with population, percent black, percent hispanic, and an urban/rural index:

\begin{align*}
&\mathbb{E}\{Y_k(\bar{0})-Y_{k-1}(\bar{0})|\bar{A}_m=(\bar{0}_{m-1},1),\bar{L}^{unemp}_m,\bar{L}^{rural}_m,\bar{L}^{pop}_m,\bar{L}^{black}_m,\bar{L}^{hisp}_m\}=\\
&\mathbb{E}\{Y_k(\bar{0})-Y_{k-1}(\bar{0})|\bar{A}_m=\bar{0},\bar{L}_m^{unemp},\bar{L}_m^{rural},\bar{L}_m^{pop},\bar{L}_m^{black},\bar{L}_m^{hisp}\}
\end{align*}
for $k>m$.

Thus, in our notation, $\bar{L}_m=(\bar{L}^{unemp}_m,\bar{L}^{rural}_m,\bar{L}^{pop}_m,\bar{L}^{black}_m,\bar{L}^{hisp}_m)$, and $\tilde{\bar{L}}_m=\bar{L}^{unemp}_m$. We estimated the parameter $\psi^*$ using a doubly robust estimator solving 
\begin{equation*}
    \mathbb{P}_N\begin{bmatrix} \phi_0(O;\psi,q_0,\hat{\pi},\hat{\mu})\\
    \phi_1(O;\psi,q_1,\hat{\pi},\hat{\mu})\end{bmatrix}=0
\end{equation*}
where $\phi_0$ and $\phi_1$ are as defined in Theorem \ref{IF_2times},  $q_0(L^{unemp}_{2014})=(1,L^{unemp}_{2014})$, and $q_1(\bar{L}^{unemp}_{2015})=(1,L^{unemp}_{2015})$. We specified and fit linear outcome regression nuisance models
\begin{align*}
\mathbb{E}(Y_{2016}-Y_{2015}|\bar{A}_{2015}=\bar{0},\bar{L}_{2015})=&\beta_1^T\{1,s(L_{2015})\}\\
\mathbb{E}(Y_{2015}-Y_{2014}|A_{2014}=0,\bar{L}_{2014})=&\beta_2^T\{1,s(L_{2014})\}\\
\mathbb{E}[\beta_1^T\{1,s(L_{2015})\}|A_{2014}=0,\bar{L}_{2014}]=&\beta_3^T\{1,s(L_{2014})\}
\end{align*}
where $s(\cdot)$ indicates cubic spline expansion and $\beta_1^T\{1,s(L_{2015})\}$ is the assumed model for $\mu_{12}$ in our notation from Section \ref{section_time_varying}. We further specified logistic regression treatment nuisance models 
\begin{align*}
    logit^{-1}\{Pr(A_{2014}=1|\bar{L}_{2014})\}=\lambda_1^T\{1,s(L_{2014})\}\\
logit^{-1}\{Pr(A_{2015}=1|A_{2014}=0,\bar{L}_{2015})\}=\lambda_2^T\{1,s(L_{2015})\}.
\end{align*}
We used the R package nleqslv \citep{dennis1996numerical} to solve all estimating equations. R code and data for the analysis are available at https://github.com/zshahn/marginal-structural-nested-mean-models-under-parallel-trends.
\begin{table}
\centering
\begin{tabular}{ |p{7.5cm}|p{4cm}|  }
 \hline
 \multicolumn{2}{|c|}{\textbf{Estimates of Medication Expansion Effects on Uninsurance Rate (pp)}} \\
 \hline
 \textbf{Estimand}& \textbf{Estimate (.95 CI)}\\
 \hline
 $\mathbb{E}\{Y_{2015}-Y_{2015}(0)|A_{2014}=1,L_{2014}^{unemp}=5\%\}$   & -4.1 (-4.5,-3.7)   \\
 \hline
 $\mathbb{E}\{Y_{2015}-Y_{2015}(0)|A_{2014}=1,L_{2014}^{unemp}=9\%\}$&   -5.6 (-6.0,-5.2)    \\
 \hline
 $\mathbb{E}\{Y_{2016}-Y_{2016}(\bar{0})|A_{2014}=1,L_{2014}^{unemp}=5\%\}$ &-4.7 (-7.2,-2.2)\\
 \hline
 $\mathbb{E}\{Y_{2016}-Y_{2016}(\bar{0})|A_{2014}=1,L_{2014}^{unemp}=9\%\}$ &-7.8 (-9.0,-6.6)\\
 \hline
 $\mathbb{E}\{Y_{2016}-Y_{2016}(\bar{0})|A_{2015}=1,L_{2015}^{unemp}=5\%\}$&   -3.3 (-4.1,-2.6) \\
 \hline
 $\mathbb{E}\{Y_{2016}-Y_{2016}(\bar{0})|A_{2015}=1,L_{2015}^{unemp}=9\%\}$&   -2.0 (-2.8,-1.2)\\
 \hline
\end{tabular}
 \caption{mSNMM estimates of heterogeneous effects of Medicaid expansion on uninsurance rates given county unemployment rate prior to expansion}
 \label{medicaid_results}
\end{table}

Table \ref{medicaid_results} summarizes selected output from the model. The results suggested that Medicaid expansion decreased uninsurance rates across years of expansion and prior year unemployment rates. The magnitude of the effect of expanding in 2014 was estimated to be larger in 2016 than in 2015, and the immediate effect of 2014 expansions was estimated to be larger than the immediate effect of the 2015 expansions. Further, the magnitude of the effect was estimated to be larger in counties that had higher recent unemployment among 2014 expanders ($\hat{\psi}_{2014,2015,1}=$-37.9 [.95 CI = (-50.1,-25.7)] and $\hat{\psi}_{2014,2016,1}=$-76.4 [.95 CI = (-166.4,13.7)]), while the reverse association was found among 2015 expanders ($\hat{\psi}_{2015,2016,1}=$33.4 [.95 CI = (4.5,62.4)]). (These coefficients are large because unemployment rate was recorded as between 0 and 1 while the uninsurance rate outcome was recorded on the percentage point scale.) We leave any interpretation of these illustrative results to social scientists. 

\section{Conclusion}
(m)SNMMs can be useful tools for estimating effects of policies and other interventions under parallel trends assumptions. In the point exposure setting, for reasons discussed in Remark \ref{remark_abadie}, our efficient and doubly robust estimator of the marginal blip model parameter has advantages over the seminal estimator of marginal effect heterogeneity proposed in \citet{abadie2005}. In the time-varying treatment setting, like other DiD methods, SNMMs can identify the effects of initiation of treatment in staggered adoption settings (see Section \ref{section_staggered}). Unlike many other alternatives, SNMMs also naturally enable the estimation of effects of repeated treatments, sustained treatment, or dynamic treatment strategies when treatment switches on and off in the observed data. Further, SNMMs can characterize heterogeneity of these effects as a function of time-varying treatment and covariate history. A drawback of `full' (non-marginal) SNMMs in practice is the need to specify a parametric blip model for the dependence of the treatment effect on all adjustment variables. Marginal SNMMs only require specification of a blip model for dependence of the treatment effect on past treatment and the history of effect modifiers of interest. This makes mSNMMs less prone to model specification than full SNMMs and also can make their results more interpretable. mSNMMs should be considered as potential analytic tools in settings where DiD approaches based on similar assumptions are currently in wide use. 

mSNMMs have attractive properties relative to other marginal modelling approaches for inferring time-varying effect modification. Unlike \textit{history-adjusted marginal structural models} \citep{van2005history}, they do not suffer from issues of incoherence \citep{robins2007invited,guo2021discussion}. This is because the marginal blip functions considered here are averages of conditional blip functions, which are variation independent (and hence coherent) when the outcome takes values on $(-\infty,\infty)$ as shown by \citet{robins2000sensitivity}. In contrast to \textit{excursion effect models} \citep{qian2021estimating}, by modelling blips based on nested sequences of covariates, we can identify $\mathbb{E}\{Y_{k}(\underbar{0})\}$ for all $k=1,...,K+1$. 


Although mSNMMs reduce the burden of blip model specification relative to full SNMMs, they do not completely eliminate it. Correct specification is still required. In settings with short time horizons and categorical effect modifiers of interest, correct blip model specification can be guaranteed through saturation. In more complex settings, misspecification can become a problem, as our blip functions model effects conditional on the entire histories of treatment and effect modifiers of interest. (Even if an investigator were only interested in effect heterogeneity as a function of the most recent covariate values, for example, they would still need to model effect heterogeneity given the entire past of those covariates and the treatment.) Forthcoming work will enable modeling effect heterogeneity conditional on relevant but non-increasing summaries of treatment and covariate history. Additionally, it might be appealing to incorporate ideas similar to \citet{lewis2020double} or \citet{jaman2024penalized} to enable regularized estimates of blip model parameters that might further reduce risk of serious misspecification under sparsity assumptions.

\section{Acknowledgements}

OD acknowledges support from the FWO grant 1222522N. ZS acknowledges support from the Interdisciplinary Research Grant program and the CUNY Office of Research... 

\bibliography{bibliography}
\newpage 
\section*{Appendix}
\appendix
\section{Identification}
\subsection{Proof of Theorem \ref{id_2tp}}
First, it follows from the identification result for the point exposure setting that $\gamma_{01}$ is identified as 
\begin{align*}
    &\gamma_{01}(\tilde{l}_0)= \mathbb{E}\{Y_1(1)-Y_1(0)|\tilde{L}_0=\tilde{l}_0,A_0=1\}\\
    &=\mathbb{E}(Y_1-Y_0|\tilde{L}_0 =\tilde{l}_0,A_0=1) - \mathbb{E}\{\mathbb{E}(Y_1-Y_0|L_0=l_0,A_0=0)|\tilde{L}_0=\tilde{l}_0,A_0=1\}.
\end{align*}

Similarly, again by the point exposure identification result, $\gamma_{12}$ is identified as
\begin{align*}
    &\gamma_{12}(\tilde{\bar{l}}_1,a_0)= \mathbb{E}\{Y_2(a_0,1)-Y_2(a_0,0)|\tilde{\bar{L}}_1=\tilde{\bar{l}}_1,\bar{A}_1=(a_0,1)\}\\
    &=\mathbb{E}(Y_2-Y_1|\tilde{\bar{L}}_1=\tilde{\bar{l}}_1,\bar{A}_1=(a_0,1)) - 
    \mathbb{E}\{\mathbb{E}(Y_2-Y_1|\bar{L}_1 =\bar{l}_1,A_0=a_0,A_1=0)|\tilde{\bar{L}}_1=\tilde{\bar{l}}_1,\bar{A}_1=(a_0,1)\}.
\end{align*}

Finally,
\begin{align*}
    \gamma_{02}(\tilde{L}_{0})=& \mathbb{E}\{Y_2(1,0)-Y_2(0,0)|\tilde{L}_0,A_0=1\}\\
    =&\mathbb{E}\{Y_2(1,0)-Y_1(0)|\tilde{L}_0,A_0=1\}-\mathbb{E}\{Y_2(0,0)-Y_1(0)|\tilde{L}_0,A_0=1\}\\
    =&\mathbb{E}[Y_2-A_1\gamma_{12}(\tilde{\bar{L}}_{1},a_0=1)-\{Y_1-\gamma_{01}(\tilde{l}_0)\}|\tilde{L}_0,A_0=1] \\
    &-\mathbb{E}[\mathbb{E}\{Y_2(1,0)-Y_1(0)|L_0,A_0=1\}|\tilde{L}_0,A_0=1]\\
    =&\mathbb{E}[Y_2-A_1\gamma_{12}(\tilde{\bar{L}}_{1},a_0=1)-\{Y_1-\gamma_{01}(\tilde{l}_0)\}|\tilde{L}_0,A_0=1] \\
    &-\mathbb{E}[\mathbb{E}\{Y_2(0,0)-Y_1(0)|L_0,A_0=0\}|\tilde{L}_0,A_0=1]\\
    =&\mathbb{E}[Y_2-A_1\gamma_{12}(\tilde{\bar{L}}_{1},a_0=1)-\{Y_1-\gamma_{01}(\tilde{l}_0)\}|\tilde{L}_0,A_0=1]  \\
    &-\mathbb{E}\left(\mathbb{E}[\mathbb{E}\{Y_2(0,0)-Y_1(0)|A_1,A_0=0,\bar{L}_1\}|L_0,A_0=0]|\tilde{L}_0,A_0=1\right)\\
    =&\mathbb{E}[Y_2-A_1\gamma_{12}(\tilde{\bar{L}}_{1},a_0=1)-\{Y_1-\gamma_{01}(\tilde{l}_0)\}|\tilde{L}_0,A_0=1] \\
    &-\mathbb{E}\left(\mathbb{E}[\mathbb{E}\{Y_2(0,0)-Y_1(0)|\bar{L}_1,A_0=0,A_1=0\}|L_0,A_0=0]|\tilde{L}_0,A_0=1\right)\\
    =&\mathbb{E}[Y_2-A_1\gamma_{12}(\tilde{\bar{L}}_{1},a_0=1)-\{Y_1-\gamma_{01}(\tilde{l}_0)\}|\tilde{L}_0,A_0=1]  \\
    &-\mathbb{E}[\mathbb{E}\{\mathbb{E}(Y_2-Y_1|\bar{L}_1,A_0=0,A_1=0)|L_0,A_0=0\}|\tilde{L}_0,A_0=1]
\end{align*}
The first equality is the definition of the blip function. The second equality comes from adding and subtracting the same quantity $\mathbb{E}\{Y_1(0)|\tilde{L}_0,A_0=1]$. The third equality follows from application of (\ref{blip_1994}) to (earlier identified) blip functions and by nested expectations. The fourth equality is by parallel trends. The fifth equality is by another application of nested expectation. The sixth equality is by another application of parallel trends. The final equality (in which all counterfactuals have been removed, establishing identification) is by consistency. 

\subsection{General time point case}
\begin{theorem}\label{id} \textbf{General Identification}\label{theorem_id_generalK} Under parallel trends Assumption \ref{parallel_trends}, $\gamma_{(k-q)k}(\tilde{\bar{L}}_{k-q},\bar{a}_{k-q-1})$ is identified as
\begin{align*}
&\gamma_{(k-q)k}(\tilde{\bar{L}}_{k-q},\bar{a}_{k-q-1})\\
&=\mathbb{E}\left[Y_k-\sum_{j=1}^{q-1}A_{k-j}\gamma_{(k-j)k}(\tilde{\bar{L}}_{k-j},\bar{A}_{k-j-1})-\left\{Y_{k-1}-\sum_{j=1}^{q-1}A_{k-j}\gamma_{(k-1-j)(k-1)}(\tilde{\bar{L}}_{k-j},\bar{A}_{k-j-1})\right\}|\tilde{\bar{L}}_{k-q},\bar{A}_{k-q}=\bar{a}_{k-q}\right]\\
&\quad -\mathbb{E}[\cdots \mathbb{E}[\mathbb{E}\{Y_k-Y_{k-1}|\bar{A}_{k-1}=(\bar{a}_{k-q-1},\underline{0}),\bar{L}_{k-1}]|\cdots]|\bar{A}_{k-q}=(\bar{a}_{k-q-1},0),\bar{L}_{k-q}]|\tilde{\bar{L}}_{k-q},\bar{A}_{k-q}=\bar{a}_{k-q}]
\end{align*}
\end{theorem}
\begin{proof} By induction in $q$ for any fixed $k$.


First, it is immediate from the point exposure identification result that $\gamma_{(k-1)k}$ is identified for arbitrary $k$ as 
\begin{align*}
    &\gamma_{(k-1)k}(\tilde{\bar{l}}_{k-1},\bar{a}_{k-2})= \mathbb{E}\{Y_k(\bar{a}_{k-2},1)-Y_k(\bar{a}_{k-2},0)|\bar{A}_{k-1}=(\bar{a}_{k-2},1),\tilde{\bar{L}}_{k-1}=\tilde{\bar{l}}_{k-1}\}\\
    &=\mathbb{E}(Y_k-Y_{k-1}|\bar{A}_{k-1}=(\bar{a}_{k-2},1),\tilde{\bar{L}}_{k-1}=\tilde{\bar{l}}_{k-1})  \\
    &\quad -\mathbb{E}\{\mathbb{E}(Y_k-Y_{k-1}|\bar{A}_{k-1}=(\bar{a}_{k-2},0),\bar{L}_{k-1}=\bar{l}_{k-1})|\bar{A}_{k-1}=(\bar{a}_{k-2},1),\tilde{\bar{L}}_{k-1}=\tilde{\bar{l}}_{k-1}\}
\end{align*}

Next, assume that we have identified all blip functions $\gamma_{(k-p)k}$ for all $p\leq q-1$ and for arbitrary $k$. We will show that the blip function $\gamma_{(k-q)k}$ is identified. The argument will proceed analogous to the $q=2$ case above. By adding and subtracting the same quantity, we will represent $\gamma_{(k-p)k}$ as a difference between conditional expectations of two differences. The first conditional expectation will be identified via blipping down by previously identified blip functions. The second conditional expectation will be identified via iteratively applying nested expectations and parallel trends until we are finally able to apply consistency.

First, we write 
\begin{align}
    \gamma_{(k-q)k}(\tilde{\bar{L}}_{k-q},\bar{a}_{k-q-1})=& \mathbb{E}\{Y_k(\bar{a}_{k-q-1},1,\underline{0})-Y_k(\bar{a}_{k-q-1},\underline{0})|\tilde{\bar{L}}_{k-q},\bar{A}_{k-q}=(\bar{a}_{k-q-1},1)\}\\
    \begin{split}\label{id_step1}
    =&\mathbb{E}\{Y_k(\bar{a}_{k-q-1},1,\underline{0})-Y_{k-1}(\bar{a}_{k-q-1},\underline{0})|\tilde{\bar{L}}_{k-q},\bar{A}_{k-q}=(\bar{a}_{k-q-1},1)\} \\
    &-\mathbb{E}\{Y_k(\bar{a}_{k-q-1},\underline{0})-Y_{k-1}(\bar{a}_{k-q-1},\underline{0})|\tilde{\bar{L}}_{k-q},\bar{A}_{k-q}=(\bar{a}_{k-q-1},1)\}.
    \end{split}
\end{align}
Consider the first term of the difference in (\ref{id_step1}). Using (\ref{blip_1994}), we can express this term as a function of observables and previously identified blip functions:
\begin{align*}
    &\mathbb{E}\{Y_k(\bar{a}_{k-q-1},1,\underline{0})-Y_{k-1}(\bar{a}_{k-q-1},\underline{0})|\tilde{\bar{L}}_{k-q},\bar{A}_{k-q}=(\bar{a}_{k-q-1},1)\}\\
    &=\mathbb{E}\Bigg[Y_k-\sum_{j=1}^{q-1}A_{k-j}\gamma_{(k-j)k}(\tilde{\bar{L}}_{k-j},\bar{A}_{k-j-1})-\\
    &\left\{Y_{k-1}-\sum_{j=1}^{q-1}A_{k-1-j}\gamma_{(k-1-j)(k-1)}(\tilde{\bar{L}}_{k-1-j},\bar{A}_{k-j-2})\right\} \,\big|\, \tilde{\bar{L}}_{k-q},\bar{A}_{k-q}=(\bar{a}_{k-q-1},1)\Bigg]
\end{align*}
Now consider the second term of the difference in (\ref{id_step1}). First, we can use nested expectations and parallel trends to expand this term to
\begin{align}
\begin{split}\label{id_step2}
    &\mathbb{E}\{Y_k(\bar{a}_{k-q-1},\underline{0})-Y_{k-1}(\bar{a}_{k-q-1},\underline{0})|\tilde{\bar{L}}_{k-q},\bar{A}_{k-q}=(\bar{a}_{k-q-1},1)\}\\
    &=\mathbb{E}[\mathbb{E}\{Y_k(\bar{a}_{k-q-1},\underline{0})-Y_{k-1}(\bar{a}_{k-q-1},\underline{0})|\bar{A}_{k-q}=(\bar{a}_{k-q-1},0),\bar{L}_{k-q}\}|\tilde{\bar{L}}_{k-q},\bar{A}_{k-q}=(\bar{a}_{k-q-1},1)].
    \end{split}
\end{align}
    Next, by iteratively applying nested expectations and parallel trends, we obtain that (\ref{id_step2}) is equal to

\begin{align}
\begin{split}
&\mathbb{E}\Bigg[\cdots\\
&\mathbb{E}[\mathbb{E}\{Y_k(\bar{a}_{k-q-1},\underline{0})-Y_{k-1}(\bar{a}_{k-q-1},\underline{0})|\bar{A}_{k-1}=(\bar{a}_{k-q-1},\underline{0}),\bar{L}_{k-1}\}|\cdots]|\bar{A}_{k-q}=(\bar{a}_{k-q-1},0),\bar{L}_{k-q})|\tilde{\bar{L}}_{k-q},\bar{A}_{k-q}=(\bar{a}_{k-q-1},0)\\
&\Bigg],
\end{split}
\end{align}
which by consistency is equal to 
\begin{align}
\begin{split}
&\mathbb{E}[\cdots \mathbb{E}[\mathbb{E}\{Y_k-Y_{k-1}|\bar{A}_{k-1}=(\bar{a}_{k-q-1},\underline{0}),\bar{L}_{k-1}]|\cdots]|\bar{A}_{k-q}=(\bar{a}_{k-q-1},0),\bar{L}_{k-q}]|\tilde{\bar{L}}_{k-q},\bar{A}_{k-q}=(\bar{a}_{k-q-1},1)],
\end{split}
\end{align}
thus eliminating counterfactuals from the expression and completing the proof by induction. 
\end{proof}

\section{Estimation - proofs of results in main texts}

\subsection{Proof of Theorem \ref{theorem:pe_IFs}}

\begin{proof}
Result (i) is a special case of the main result in Theorem \ref{theorem:if_gen}.

Result (ii) follows by application of the generalized information matrix equality. Specifically, 
following Theorem 5.3 in  \citet{newey1994large}, the optimal index function $q_{eff}(\tilde{L}_0)$ satisfies
\[-\frac{\partial}{\partial \psi} \mathbb{E}\left\{\phi(O;\psi,q,\pi^*,\mu^*)\right\}|_{\psi=\psi^*}=\mathbb{E}\left\{\phi(O;\psi^*,q,\pi^*,\mu^*)\phi(O;\psi^*,q_{eff},\pi^*,\mu^*)^T\right\}\]
Because this holds for arbitrary $q$, it implies in particular, taking $q(\tilde{L}_0)=I(\tilde{L}_0=\tilde{l}_0)$,  that 
\begin{align*}
\mathbb{E}\left\{A\frac{\partial}{\partial \psi}\gamma(\tilde{L}_0;\psi)|_{\psi=\psi^*}|\tilde{L}_0\right\}&=\mathbb{E}\left(\left[\frac{A-\pi^*(L_0)}{1-\pi^*(L_0)}\{Y_1-Y_0-\mu^*(L_0)\}-A\gamma(\tilde{L}_0;\psi^*)\right]^2|\tilde{L}_0\right)q_{eff}(\tilde{L}_0)^T
\end{align*}
since $\gamma(\cdot;\cdot)$ is assumed to be smooth in $\psi$, and hence 
\begin{align*}
\mathbb{E}(A|\tilde{L}_0)\frac{\partial}{\partial \psi}\gamma(\tilde{L}_0;\psi)|_{\psi=\psi^*}&= \textrm{var}\left[\frac{A-\pi^*(L_0)}{1-\pi^*(L_0)}\{Y_1-Y_0-\mu^*(L_0)\}-A\gamma(\tilde{L}_0;\psi^*)|\tilde{L}_0\right]q_{eff}(\tilde{L}_0)^T.
\end{align*}
The result then follows by rearrangement of the above.

To show result (iii), recall that $\psi^*$ solves
\begin{align}\label{equation:key_res}
\mathbb{E}[q(\tilde{L}_0)\{Y_1-Y_0-\mu^*(L_0)-A\gamma(\tilde{L}_0;\psi^*)\}]=0
\end{align}
Note also that $\phi(O;\psi^*,q,\pi^\dagger,\mu^\dagger)$ can be re-written as 
\begin{align*}
q(\tilde{L}_0)\left[\left\{1-\frac{1-A}{1-\pi^\dagger(L_0)}\right\}\left\{Y_1-Y_0-\mu^\dagger(L_0)\right\}-A\gamma(\tilde{L}_0;\psi^*)\right].
\end{align*}
Suppose first that $\mu^\dagger(L_0)=\mu^*(L_0)$  but $\pi^\dagger(L_0)\neq \pi(L_0)$. Then
\begin{align*}
&\mathbb{E}\left(q(\tilde{L}_0)\left[\left\{1-\frac{1-A}{1-\pi^\dagger(L_0)}\right\}\left\{Y_1-Y_0-\mu^*(L_0)\right\}-A\gamma(\tilde{L}_0;\psi^*)\right]\right)\\
&=\mathbb{E}[q(\tilde{L}_0)\{Y_1-Y_0-\mu^*(L_0)-A\gamma(\tilde{L}_0;\psi^*)\}]-\mathbb{E}\left[q(\tilde{L}_0) \frac{(1-A)}{\{1-\pi^\dagger(L_0)\}}\left\{Y_1-Y_0-\mu^*(L_0)\right\}\right]\\
&=-\mathbb{E}\left[q(\tilde{L}_0) \frac{(1-A)}{\{1-\pi^\dagger(L_0)\}}\left\{Y_1-Y_0-\mu^*(L_0)\right\}\right] &\textrm{\{By \eqref{equation:key_res}\}}\\
&=-\mathbb{E}\left[q(\tilde{L}_0) \frac{(1-A)}{\{1-\pi^\dagger(L_0)\}}\left\{\mu^*(L_0)-\mu^*(L_0)\right\}\right] &\textrm{(Iterated expectation)}\\
&=0
\end{align*}

Alternatively,  suppose that $\mu^\dagger(L_0)\neq \mu^*(L_0)$  but $\pi^\dagger(L_0)= \pi^*(L_0)$.  Then
\begin{align*}
&\mathbb{E}\left(q(\tilde{L}_0)\left[\left\{1-\frac{1-A}{\{1-\pi^*(L_0)\}}\right\}\left\{Y_1-Y_0-\mu^\dagger(L_0)\right\}-A\gamma(\tilde{L}_0;\psi^*)\right]\right)\\
&=\mathbb{E}[q(\tilde{L}_0)\{Y_1-Y_0-\mu^\dagger(L_0)-A\gamma(\tilde{L}_0;\psi^*)\}] -\mathbb{E}\left[q(\tilde{L}_0) \frac{\mathbb{E}(1-A|L_0)}{\{1-\pi^*(L_0)\}}\left\{\mu^*(L_0)-\mu^\dagger(L_0)\right\}\right] &\textrm{(Iterated expectation)}\\
&=\mathbb{E}[q(\tilde{L}_0)\{Y_1-Y_0-\mu^\dagger(L_0)-A\gamma(\tilde{L}_0;\psi^*)\}]-\mathbb{E}[q(\tilde{L}_0)\left\{\mu^*(L_0)-\mu^\dagger(L_0)\right\}]\\
&=\mathbb{E}[q(\tilde{L}_0)\{Y_1-Y_0-\mu^*(L_0)-A\gamma(\tilde{L}_0;\psi^*)\}] \\
&=0 &\textrm{\{By \eqref{equation:key_res}\}}.
\end{align*}
\end{proof}

\subsection{Proof of Theorem \ref{theorem:eff_score_2t}}

\begin{proof}
It follows from Theorems \ref{theorem:estimation_2t} and \ref{theorem:if_gen} that the set of influence functions for $\psi^*$under $\mathcal{M}$ (up to scaling factors) can be decomposed as $\mathcal{C}_0\cup \mathcal{C}_1$, where
\begin{align*}
\mathcal{C}_0 &\equiv \left\{q_{0}(\tilde{L}_0)\epsilon_0(O;\psi^*,\pi^*,\mu^*):q_0\textrm{ unrestricted}\right\}\cap \mathcal{L}_2\\
\mathcal{C}_1 &\equiv \left\{q_{1}(\tilde{\bar{L}}_1,A_0)\epsilon_1(O;\psi^*,\pi^*,\mu^*):q_1\textrm{ unrestricted}\right\}\cap \mathcal{L}_2
\end{align*}
Then to derive the efficient score, we will orthogonalize $\mathcal{C}_0\cup \mathcal{C}_1$ as $\mathcal{C}^*_0+\mathcal{C}_1$ where $\mathcal{C}^*_0\equiv \Pi(C_0|C_1^\perp)$, $\Pi(\cdot|\cdot)$ is the $L_2$ orthogonal projection operator and $\perp$ denotes the orthogonal complement. One can show that 
\begin{align*}
\mathcal{C}_0^*=\left\{\mathfrak{q}_{0}(\tilde{L}_0)\mathfrak{e}_0(O;\psi,\pi^*,\mu^*):\mathfrak{q}_{0}\textrm{ unrestricted}\right\}\cap \mathcal{L}_2
\end{align*}
noting that $\mathbb{E}\{\mathfrak{e}_0(O;\psi,\pi^*,\mu^*)|\tilde{\bar{L}}_1,A_0\}=0$ and $\mathbb{E}\{\mathfrak{e}_0(O;\psi,\pi^*,\mu^*)|\tilde{L}_0\}=0$. Then the result follows from an application of the generalized information matrix equality as in the proof of Theorem \ref{theorem:pe_IFs}.
\end{proof}

\section{Estimation - the general case}

\subsection{The set of influence functions}

 For each $m=0,...,K$ and $k=m+1,...,K+1$, we assume that for each possible value $\tilde{\bar{l}}_m$ of $\tilde{\bar{L}}_m$, there exists a known function $\gamma_{mk}(\tilde{\bar{l}}_m;\cdot):\Psi \mapsto \mathbb{R}$, $0\in \Psi \subset \mathbb{R}^p$, satisfying $\gamma_{mk}(\tilde{\bar{l}}_m;0)=0$, such that 
 \begin{align*}
 \gamma_{mk}(\tilde{\bar{l}}_m)=\gamma_{mk}(\tilde{\bar{l}}_m;\psi^*)
 \end{align*}
 where $\psi^*\in \Psi$ is unknown. For $m=0,...,K$ and $k=m+1,...,K+1$, we recursively define
\begin{align*}
\mu^*_{kk}&\equiv Y_k-Y_{k-1} \\
\mu^*_{k-1,k}(\bar{L}_{k-1},\bar{A}_{k-2})&\equiv \mathbb{E}(\mu^*_{kk}|\bar{L}_{k-1},\bar{A}_{k-2},A_{k-1}=0)
\end{align*}
so that for $j=k-2,...,m$
\begin{align*}
\mu^*_{jk}(\bar{L}_{j},\bar{A}_{j-1})&\equiv \mathbb{E}(\mu^*_{j+1,k}|\bar{L}_{j},\bar{A}_{j-1},A_{j}=0).
\end{align*}
We let $U_m(\psi^*)=\{U_{m,m+1}(\psi^*),U_{m,m+2}(\psi^*),...,U_{m,K+1}(\psi^*)\}^T$ denote a $(K+1-m)$-length vector, with
\begin{align*}
U_{mk}(\psi)\equiv &Y_k-\sum_{j=m}^{k-1}A_{j}\gamma_{jk}(\tilde{\bar{L}}_j,\bar{A}_{j-1};\psi^*)-\left\{Y_{k-1}-\sum_{j=m}^{k-1}A_{j}\gamma_{j,k-1}(\tilde{\bar{L}}_j,\bar{A}_{j-1};\psi^*)\right\}
\end{align*}
for $k=m+1,...,K+1$, where $\gamma_{m,m}(\tilde{\bar{L}}_m,\bar{A}_{m-1};\psi^*)=0$.\\ Similarly, $\dot{U}_{m}(\psi^*,\mu^*)\equiv \{\dot{U}_{m,m+1}(\psi^*,\mu^*_{m,m+1}),..,\dot{U}_{m,K+1}(\psi^*,\mu^*_{m,K+1})\}^T$ is a $(K+1-m)$-length vector with
\begin{align*}
\dot{U}_{mk}(\psi^*,\mu^*_{mk})\equiv&U_{mk}(\psi^*)-\mu^*_{mk}(\bar{L}_m,\bar{A}_{m-1})
\end{align*}
for $k=m+1,...,K+1$. We define the model $\mathcal{M}$ to denote the collection of distributions $P$ for which
\begin{align*}
0=\mathbb{E}_P\{\dot{U}_{{mk}}(\psi,\mu_{{mk}_P})|\tilde{\bar{L}}_m,\bar{A}_{m-1}\}\end{align*}
for some $\psi \in \Psi$; here, $\mu_{{mk}_P}$ refers to the nested iterated conditional expectation function (as well as all expectations within) evaluated at $P$. We assume each blip function is smooth in $\psi$.

Then the following result describes the class of influence functions in the general setting.
\begin{theorem}\label{theorem:if_gen}
The set of influence functions for all RAL estimators of $\psi^*$ under $\mathcal{M}$ is given by
\[\left\{D(\psi^*,q)^{-1}\sum^K_{m=0}\phi_{m}(O;\psi^*,q_m,\pi^*,\mu^*) :q_0,...,q_k\textrm{ unrestricted}\right\}\cap \mathcal{L}_2
\]
where for any $\psi\in\Psi$ and any arbitrary functions $\pi_m(\bar{L}_m)$ and $\mu_{mk}(\bar{L}_m)$ for $m=0,...,K$ and $k=m+1,...,K+1$ (such that $\pi_m(\bar{L}_m)\neq1$ for all $m$ with probability one)
\begin{align*}
\phi_m(O;\psi,q_m,\pi,\mu) \equiv & q_{m}(\tilde{\bar{L}}_m,\bar{A}_{m-1})\epsilon_m(O;\psi,q_m,\pi,\mu),\\
D(\psi,q) \equiv& -\frac{\partial}{\partial \psi}\mathbb{E}\left\{\sum^K_{m=0} q_{m}(\tilde{\bar{L}}_m,\bar{A}_{m-1})U_{m}(\psi)\right\}|_{\psi=\psi}.
\end{align*}
Moreover, $q_{m}(\tilde{\bar{L}}_m,\bar{A}_{m-1})$ denotes a $p \times (K+1-m)$-dimensional index function such that $D(\psi^*,q)^{-1}$ exists and $\epsilon_{m}(\psi,\pi,\mu)$ is a $(K+1-m)$-length vector with $k$-th element
\begin{align*}
\epsilon_{mk}(\psi,\pi,\mu)&\equiv\dot{U}_{{mk}}(\psi,\mu_{mk})-\sum^{k-1}_{j=m} \left\{ \prod^j_{r=m}\frac{(1-A_r)}{\{1-\pi_r(\bar{L}_r,\bar{A}_{r-1})\}}\right\}\{\mu_{j+1,k}(\bar{L}_{j+1},\bar{A}_{j})-\mu_{jk}(\bar{L}_{j},\bar{A}_{j-1})\}.
\end{align*}
\end{theorem}

\begin{proof}
Note that $\psi^*$ satisfies
\begin{align*}
0=\mathbb{E}\{q_{mk}(\tilde{\bar{L}}_m,\bar{A}_{m-1})\dot{U}_{mk}(\psi^*,\mu^*_{mk})\}
\end{align*}
$\forall m=0,...,K$ and $k=m+1,...,K+1$, where $q_{mk}(\tilde{\bar{L}}_m,\bar{A}_{m-1})$ is an arbitrary  scalar function of $(\tilde{\bar{L}}_m,\bar{A}_{m-1})$. Furthermore, it follows that 
\begin{align}\label{equation:pwd_moment}
0=\frac{\partial }{\partial t} \mathbb{E}_t\{q_{mk}(\tilde{\bar{L}}_m,\bar{A}_{m-1})\dot{U}_{{mk}}(\psi_t,\mu_{{mk}_t})\}|_{t=0}
\end{align}
$\forall m=0,...,K$ and $k=m+1,...,K+1$, where we index by the subscript $t$ to denote that the function or functional is evaluated at the law $P_t$. We use the shorthand $\mathbb{E}_{t}(\cdot)$ for $\mathbb{E}_{P_t}(\cdot)$ and $t=0$ at the true law $\mathbb{P}$. 
 
Evaluating \eqref{equation:pwd_moment} at a given $m$ and $k$ gives
\begin{align*}
&\frac{\partial }{\partial t} \mathbb{E}_t\{q_{mk}(\tilde{\bar{L}}_m,\bar{A}_{m-1})\dot{U}_{{mk}}(\psi_t,\mu_{{mk}_t})\}|_{t=0}\\
&=\frac{\partial }{\partial t} \mathbb{E}_t\{q_{mk}(\tilde{\bar{L}}_m,\bar{A}_{m-1})\dot{U}_{{mk}}(\psi^*,\mu^*_{mk})\}|_{t=0}+\mathbb{E}\left\{q_{mk}(\tilde{\bar{L}}_m,\bar{A}_{m-1})\frac{\partial }{\partial t} \dot{U}_{{mk}}(\psi_t,\mu_{{mk}_t})|_{t=0}\right\}\\
&=\frac{\partial }{\partial t} \mathbb{E}_t\{q_{mk}(\tilde{\bar{L}}_m,\bar{A}_{m-1})\dot{U}_{{mk}}(\psi^*,\mu^*_{mk})\}|_{t=0}
+\mathbb{E}\left\{q_{mk}(\tilde{\bar{L}}_m,\bar{A}_{m-1})\frac{\partial }{\partial t} U_{{mk}}(\psi_t)|_{t=0}\right\}\\
&\quad-\mathbb{E}\left\{q_{mk}(\tilde{\bar{L}}_m,\bar{A}_{m-1})\frac{\partial }{\partial t}\mu_{{mk}_t}(\bar{L}_m,\bar{A}_{m-1})|_{t=0}\right\}.
\end{align*}
If we consider each of the terms on the right hand side of the final equality, then firstly
\begin{align*}
&\frac{\partial }{\partial t} \mathbb{E}_t\{q_{mk}(\tilde{\bar{L}}_m,\bar{A}_{m-1})\dot{U}_{{mk}}(\psi^*,\mu^*_{mk})\}|_{t=0}\\
&=\mathbb{E}\{q_{mk}(\tilde{\bar{L}}_m,\bar{A}_{m-1})\dot{U}_{{mk}}(\psi^*,\mu^*_{{mk}})S(Y_{k},Y_{k-1},\bar{L}_{k-1},\bar{A}_{k-1})^T\}\\
&=\mathbb{E}\left(q_{mk}(\tilde{\bar{L}}_m,\bar{A}_{m-1})\dot{U}_{{mk}}(\psi^*,\mu^*_{{mk}}) [\mathbb{E}\{S(O)|Y_{k},Y_{k-1},\bar{L}_{k-1},\bar{A}_{k-1}\}^T-\mathbb{E}\{S(O)\}^T]\right)\\
&=\mathbb{E}[q_{mk}(\tilde{\bar{L}}_m,\bar{A}_{m-1})\dot{U}_{{mk}}(\psi^*,\mu^*_{{mk}})\mathbb{E}\{S(O)|Y_{k},Y_{k-1},\bar{L}_{k-1},\bar{A}_{k-1}\}^T]\\
&=\mathbb{E}\{q_{mk}(\tilde{\bar{L}}_m,\bar{A}_{m-1})\dot{U}_{{mk}}(\psi^*,\mu^*_{{mk}})S(O)^T\}.
\end{align*}
where $S(O)$ is the score function. Secondly, by application of the chain rule,
\begin{align*}
&\mathbb{E}\left\{q_{mk}(\tilde{\bar{L}}_m,\bar{A}_{m-1})\frac{\partial }{\partial t} U_{{mk}}(\psi_t)|_{t=0}\right\}\\
&=\mathbb{E}\left\{q_{mk}(\tilde{\bar{L}}_m,\bar{A}_{m-1})\frac{\partial }{\partial \psi} U_{mk}(\psi)|_{\psi=\psi^*}\right\} \frac{\partial \psi_t}{\partial t}|_{t=0}.
\end{align*}
Finally, with abuse of notation, 
\begin{align*}
&\mathbb{E}\left\{q_{mk}(\tilde{\bar{L}}_m,\bar{A}_{m-1})\frac{\partial }{\partial t}\mu_{{mk}_t}(\bar{L}_{m},\bar{A}_{m-1})|_{t=0}\right\}\\
&=\mathbb{E}\left(q_{mk}(\tilde{\bar{L}}_m,\bar{A}_{m-1}) \left[\sum^{k-1}_{j=m} \left\{ \prod^j_{r=m}\frac{(1-A_r)}{\{1-\pi^*_r(\bar{L}_r,\bar{A}_{r-1})\}}\right\}\mu^*_{j+1,k}(\bar{L}_{j+1},\bar{A}_{j})S(L_{j+1}|\bar{L}_{j},\bar{A}_{j})^T \right]\right)\\
&=\mathbb{E}\left\{q_{mk}(\tilde{\bar{L}}_m,\bar{A}_{m-1}) \left(\sum^{k-1}_{j=m} \left\{ \prod^j_{r=m}\frac{(1-A_r)}{\{1-\pi^*_r(\bar{L}_r,\bar{A}_{r-1})\}}\right\}\mu^*_{j+1,k}(\bar{L}_{j+1},\bar{A}_{j})[\mathbb{E}\{S(O)|\bar{L}_{j+1},\bar{A}_j\}^T-\mathbb{E}\{S(O)|\bar{L}_{j},\bar{A}_j\}^T] \right)\right\}\\
&=\mathbb{E}\left(q_{mk}(\tilde{\bar{L}}_m,\bar{A}_{m-1}) \left[\sum^{k-1}_{j=m} \left\{ \prod^j_{r=m}\frac{(1-A_r)}{\{1-\pi^*_r(\bar{L}_r,\bar{A}_{r-1})\}}\right\}\{\mu^*_{j+1,k}(\bar{L}_{j+1},\bar{A}_{j})-\mu^*_{jk}(\bar{L}_{j},\bar{A}_{j-1})\} \right]S(O)^T\right)
\end{align*}
where the second and third equalities can be shown using a reverse induction argument; see e.g. the proof of Theorem 1 of \citet{van2012targeted}. 
\[\]

Putting this together, we have 
\begin{align*}
&\frac{\partial }{\partial t} \mathbb{E}_t\{q_{mk}(\tilde{\bar{L}}_m,\bar{A}_{m-1})\dot{U}_{{mk}}(\psi_t,\mu_{{mk}_t})\}|_{t=0}\\
&=\mathbb{E}\left(q_{mk}(\tilde{\bar{L}}_m,\bar{A}_{m-1}) \left[\dot{U}_{{mk}}(\psi^*)-\sum^{k-1}_{j=m} \left\{ \prod^j_{r=m}\frac{(1-A_r)}{\{1-\pi^*_r(\bar{L}_r,\bar{A}_{r-1})\}}\right\}\{\mu^*_{j+1,k}(\bar{L}_{j+1},\bar{A}_{j})-\mu^*_{jk}(\bar{L}_{j},\bar{A}_{j-1})\}\right]S(O)^T\right) \\
&\quad+\mathbb{E}\left\{q_{mk}(\tilde{\bar{L}}_m,\bar{A}_{m-1})\frac{\partial }{\partial \psi} U_{mk}(\psi)|_{\psi=\psi^*}\right\} \frac{\partial \psi_t}{\partial t}|_{t=0}
\end{align*}

Since the previous arguments hold for all $m=0,...,K$ and $k=m+1,...,K+1$, by combining all results across time points, it follows that:
\begin{align*}
0=\mathbb{E}\left\{ \sum^{K}_{m=0}\phi_{m}(O;\psi^*,q_m,\pi^*,\mu^*)S(O)^T\right\}+\frac{\partial}{\partial \psi}\mathbb{E}\left\{\sum^{K}_{m=0}q_{m}(\tilde{\bar{L}}_m,\bar{A}_{m-1})U_{m}(\psi)\right\}|_{\psi=\psi^*} \frac{\partial \psi_t}{\partial t}|_{t=0}.
\end{align*}
Since $q_{m}(\tilde{\bar{L}}_m,\bar{A}_{m-1})$ is a conformable $p\times 1$ function, we can rearrange the above such that
\begin{align*}
\frac{\partial \psi_t}{\partial t}|_{t=0}= D(\psi^*,q)^{-1}\mathbb{E}\left\{\sum^K_{m=0}\phi_{m}(O;\psi^*,q_m,\pi^*,\mu^*)S(O)^T\right\}
\end{align*}
Then the result immediately follows. 
\end{proof}

\subsection{The cross-fit estimator $\hat{\psi}^{cf}$}

\begin{algorithm}
\caption{Implementation of a cross-fit DR-ML estimator: time-varying setting}\label{algorithm:cfdrml-gen}
 Randomly split the $N$ units into $S$ disjoint folds of size $n_s\approx N/S$.\\
For each $s=1,...,S$, construct estimators $\hat{\mu}^{(-s)}$ of $\mu^*$ and $\hat{\pi}^{(-s)}$ of $\pi^*$, using all the units except those in fold $s$.\\
For each $s=1,...,S$, compute $\hat{\psi}^{(s)}$ from the $s$-th fold as the (assumed unique) solution to the estimating equations $\mathbb{P}_{n_s}\{\sum^K_{m=0}\phi_m(O;\psi,q_m,\hat{\pi}^{(-1)},\hat{\mu}^{(-1)})\}=0$ where $\mathbb{P}_{n_s}$ refers here to the empirical measure in the $s$-th fold.\\
Take the weighted average of the $S$ estimators:  $\hat{\psi}^{cf}=\sum^{S}_{s=1}\left(\frac{n_s}{N}\right)\hat{\psi}^{(s)}$.
\end{algorithm}

\subsection{Asymptotic linearity of the cross-fit estimator}


\begin{theorem}\label{estimation_theorem_2t}
Suppose the following conditions hold:
\begin{enumerate}
    \item The map $\psi \mapsto \mathbb{P}\left\{\sum^K_{m=0}\phi_m(o;\psi,q_0,\pi,\mu)\right\}$ is differentiable at $\psi^*$ uniformly in $(\pi,\mu)$. 
    \item The matrix $D(\psi^*,q)$ is nonsingular.
    \item For $m=0,...,K$ and  and $s=1,...,S$, $\{1-\hat{\pi}^{(-s)}_m(\bar{L}_m,\bar{A}_{m-1})\}\geq c>0$ and $\{1-\pi^*_m(\bar{L}_m,\bar{A}_{m-1})\}\geq c>0$ with probability one for a constant $c$.
    \item For $m=0,...,K$ and for a chosen $p \times (K+1-m)$ function $q_m(\tilde{\bar{L}}_m,\bar{A}_{m-1})$, the maximum absolute value of any element in $q_m(\tilde{\bar{L}}_m,\bar{A}_{m-1})$ is upper bounded by $C$ with probability one, where $C<\infty$.
    \item $\hat{\psi}^{(s)}-\psi^*=o_{\mathbb{P}}(1)$, $\norm{\hat{\pi}^{(-s)}_{m} - \pi_{m}^*}=o_\mathbb{P}(1)$ and $\norm{\hat{\mu}^{(-s)}_{mk}-\mu_{mk}^*}=o_\mathbb{P}(1)$ for $m=0,...,K$, $k=m+1,...,K+1$ and $s=1,...,S$.
    \item For $m=0,...,K$, $k=m+1,...,K+1$ and $s=1,...,S$,
    \begin{align*}
    \norm{\hat{\pi}^{(-s)}_{m} - \pi_{m}^*}\norm{\hat{\mu}^{(-s)}_{mk}-\mu_{mk}^*}=& o_\mathbb{P}(N^{-1/2}).
    \end{align*} 
\item For any fixed ($\pi,\mu$), the class $\left\{\sum^K_{m=0}\phi_m(o;\psi,q_0,\pi,\mu):\psi\in \Psi\right\}$ is Donsker in $\psi$. 
\end{enumerate}
Then $\hat{\psi}^{cf}$ satisfies the expansion
\begin{equation*}
\sqrt{N}(\hat{\psi}^{cf}-\psi^{*})=D(\psi^*,q)^{-1} \frac{1}{\sqrt{N}}\sum_{i=1}^N\sum_{m=0}^K\phi_m(O_i;\psi^*,q_m,\pi^*,\mu^*)+o_\mathbb{P}(1)
\end{equation*}
\end{theorem}

\begin{proof}
By application of Lemma \ref{lemma:cf}, it follows that 
\begin{align*}
\sqrt{N}(\hat{\psi}^{cf}-\psi^{*})=&D(\psi^*,q)^{-1}\sqrt{N}(\mathbb{P}_N-\mathbb{P})\left\{\sum_{m=0}^K\phi_m(\psi^*,q_m,\pi^*,\mu^*)\right\}\\&+\sqrt{N}O_\mathbb{P}\left(\sum^{S}_{s=1}\left(\frac{n_s}{N}\right)\mathbb{P}\left\{\sum_{m=0}^K \phi_m(\psi^*,q_m,\hat{\pi}^{(-s)},\hat{\mu}^{(-s)})\right\}\right)+o_\mathbb{P}\left(1\right)
\end{align*}
Since the number of folds is fixed, the main result follows if we can show that
\begin{align}\label{eq:target_res}
\mathbb{P}\left\{\sum_{m=0}^K \phi_m(\psi^*,q_m,\hat{\pi}^{(-s)},\hat{\mu}^{(-s)})\right\}=o_\mathbb{P}(N^{-1/2})
\end{align}
for $s=1,...,S$. 

Since a random vector converges in probability iff each component of the vector converges to its corresponding limit, we will proceed by showing that each component of the vector on the left hand side of \eqref{eq:target_res} converges to zero sufficiently fast.  For any $m=0,...K$ and $s=1,...,S$, by definition we have 
\begin{align*}
\sum^K_{m=0}\phi_m(O;\psi^*,q_m,\hat{\pi}^{(-s)},\hat{\mu}^{(-s)})=\begin{pmatrix}
\sum^{K}_{m=0}\sum^{K+1}_{k=m+1}q_{mk1}(\tilde{\bar{L}}_m,\bar{A}_{m-1})\epsilon_{mk}(O;\psi^*,\hat{\pi}^{(-s)},\hat{\mu}^{(-s)})\\
\vdots \\
\sum^{K}_{m=0}\sum^{K+1}_{k=m+1}q_{mkp}(\tilde{\bar{L}}_m,\bar{A}_{m-1})\epsilon_{mk}(O;\psi^*,\hat{\pi}^{(-s)},\hat{\mu}^{(-s)})
\end{pmatrix}
\end{align*}
where $q_{mkj}(\tilde{\bar{L}}_m,\bar{A}_{m-1})$ is the  $(j,k)$th element of $q_m(\tilde{\bar{L}}_m,\bar{A}_{m-1})$. 

Define $\mathds{I}_{mk}=I(k>m+1)$.  Now for any $j=1,...p$,
\begin{align*}
&\bigg|\mathbb{P}\left\{\sum^{K}_{m=0}\sum^{K+1}_{k=m+1}q_{mkj}\epsilon_{mk}(\psi^*,\hat{\pi}^{(-s)},\hat{\mu}^{(-s)})\right\}\bigg|&\\
&= \Bigg|\mathbb{P}\left(\sum^{K}_{m=0}\sum^{K+1}_{k=m+1}q_{mkj}\left[\frac{(\pi_m^*-\hat{\pi}^{(-s)}_m)}{(1-\hat{\pi}^{(-s)}_m)}(\mu^*_{m,k}-\hat{\mu}^{(-s)}_{m,k})+\mathds{I}_{mk}\sum^{k-1}_{j=m+1} \left\{ \prod^{j-1}_{r=m}\frac{(1-A_r)}{(1-\hat{\pi}^{(-s)}_r)}\right\}\frac{(\pi^*_{j}-\hat{\pi}^{(-s)}_{j})}{(1-\hat{\pi}^{(-s)}_{j})}(\mu^*_{jk}-\hat{\mu}^{(-s)}_{jk})\right]\right)\Bigg| \\
&\leq \mathbb{P}\left(\Bigg|\sum^{K}_{m=0}\sum^{K+1}_{k=m+1}q_{mkj}\left[\frac{(\pi_m^*-\hat{\pi}^{(-s)}_m)}{(1-\hat{\pi}^{(-s)}_m)}(\mu^*_{m,k}-\hat{\mu}^{(-s)}_{m,k})+\mathds{I}_{mk}\sum^{k-1}_{j=m+1} \left\{ \prod^{j-1}_{r=m}\frac{(1-A_r)}{(1-\hat{\pi}^{(-s)}_r)}\right\}\frac{(\pi^*_{j}-\hat{\pi}^{(-s)}_{j})}{(1-\hat{\pi}^{(-s)}_{j})}(\mu^*_{jk}-\hat{\mu}^{(-s)}_{jk})\right]\Bigg|\right) \\
&\leq \mathbb{P}\left(\sum^{K}_{m=0}\sum^{K+1}_{k=m+1}\Bigg|q_{mkj}\left[\frac{(\pi_m^*-\hat{\pi}^{(-s)}_m)}{(1-\hat{\pi}^{(-s)}_m)}(\mu^*_{m,k}-\hat{\mu}^{(-s)}_{m,k})+\mathds{I}_{mk}\sum^{k-1}_{j=m+1} \left\{ \prod^{j-1}_{r=m}\frac{(1-A_r)}{(1-\hat{\pi}^{(-s)}_r)}\right\}\frac{(\pi^*_{j}-\hat{\pi}^{(-s)}_{j})}{(1-\hat{\pi}^{(-s)}_{j})}(\mu^*_{jk}-\hat{\mu}^{(-s)}_{jk})\right]\Bigg|\right) \\
&= \mathbb{P}\left(\sum^{K}_{m=0}\sum^{K+1}_{k=m+1}|q_{mkj}|\Bigg|\frac{(\pi_m^*-\hat{\pi}^{(-s)}_m)}{(1-\hat{\pi}^{(-s)}_m)}(\mu^*_{m,k}-\hat{\mu}^{(-s)}_{m,k})+\mathds{I}_{mk}\sum^{k-1}_{j=m+1} \left\{ \prod^{j-1}_{r=m}\frac{(1-A_r)}{(1-\hat{\pi}^{(-s)}_r)}\right\}\frac{(\pi^*_{j}-\hat{\pi}^{(-s)}_{j})}{(1-\hat{\pi}^{(-s)}_{j})}(\mu^*_{jk}-\hat{\mu}^{(-s)}_{jk})\Bigg|\right) \\
&\leq C \mathbb{P}\left(\sum^{K}_{m=0}\sum^{K+1}_{k=m+1}\Bigg|\frac{(\pi_m^*-\hat{\pi}^{(-s)}_m)}{(1-\hat{\pi}^{(-s)}_m)}(\mu^*_{m,k}-\hat{\mu}^{(-s)}_{m,k})+\mathds{I}_{mk}\sum^{k-1}_{j=m+1} \left\{ \prod^{j-1}_{r=m}\frac{(1-A_r)}{(1-\hat{\pi}^{(-s)}_r)}\right\}\frac{(\pi^*_{j}-\hat{\pi}^{(-s)}_{j})}{(1-\hat{\pi}^{(-s)}_{j})}(\mu^*_{jk}-\hat{\mu}^{(-s)}_{jk})\Bigg|\right) \\
&=C\sum^{K}_{m=0}\sum^{K+1}_{k=m+1}\mathbb{P}\left(\Bigg|\frac{(\pi_m^*-\hat{\pi}^{(-s)}_m)}{(1-\hat{\pi}^{(-s)}_m)}(\mu^*_{m,k}-\hat{\mu}^{(-s)}_{m,k})+\mathds{I}_{mk}\sum^{k-1}_{j=m+1} \left\{ \prod^{j-1}_{r=m}\frac{(1-A_r)}{(1-\hat{\pi}^{(-s)}_r)}\right\}\frac{(\pi^*_{j}-\hat{\pi}^{(-s)}_{j})}{(1-\hat{\pi}^{(-s)}_{j})}(\mu^*_{jk}-\hat{\mu}^{(-s)}_{jk})\Bigg|\right) 
\end{align*}
where the first step follows from Lemma \ref{lemma:drift}, the second and the third from the triangle inequality and the fifth due to Condition 4.
We repeatedly use the elementary inequality $a+b\leq c+d$ if $a\leq c$ and $b\leq d$. 

Now 
\begin{align*}
&\sum^{K}_{m=0}\sum^{K+1}_{k=m+1}\mathbb{P}\left(\Bigg|\frac{(\pi_m^*-\hat{\pi}^{(-s)}_m)}{(1-\hat{\pi}^{(-s)}_m)}(\mu^*_{m,k}-\hat{\mu}^{(-s)}_{m,k})+\mathds{I}_{mk}\sum^{k-1}_{j=m+1} \left\{ \prod^{j-1}_{r=m}\frac{(1-A_r)}{(1-\hat{\pi}^{(-s)}_r)}\right\}\frac{(\pi^*_{j}-\hat{\pi}^{(-s)}_{j})}{(1-\hat{\pi}^{(-s)}_{j})}(\mu^*_{jk}-\hat{\mu}^{(-s)}_{jk})\Bigg|\right) \\
&=O_{\mathbb{P}}\left(\sum_{m=0}^{K} \sum_{k=m+1}^{K+1}\norm{\hat{\pi}_{m} - \pi_{m}^*}\norm{\hat{\mu}_{mk}-\mu_{mk}^*}\right)\\
&=o_{\mathbb{P}}(N^{-1/2})
\end{align*}
where the first equality follows from Lemma \ref{lemma:product_rate} and the second from Condition 6. Then this implies \eqref{eq:target_res}, since $p$ is fixed and $s$ is arbitrary.
\end{proof}

\section{Alternative Derivation Referred to in Remark 2}\label{section_appendix_alternate}
Let $X_0\equiv L_0 \backslash \tilde{L}_0$; then define a density $\tilde{f}(O)$ which differs from the observed data density in two respects. Firstly, $\tilde{f}(A|\tilde{L}_0)$ is substituted for $f(A|L_0)$, where $\tilde{f}(A|\tilde{L}_0)$ is a user-specified density which is absolutely continuous with respect to $f(A|L_0)$. Secondly, $f(X_0|\tilde{L}_0)$ is replaced by $f(X_0|A=1,\tilde{L}_0)$; the observed conditional distribution of $X_0$ is switched to the observed conditional distribution \textit{in the treated}. Note that under $\tilde{f}(O)$, $A$ is conditionally independent of $X_0$ given $\tilde{L}_0$ (e.g. there is no confounding by $X_0$). Let $\tilde{\mathbb{E}}(\cdot)$ refer to an expectation taken with respect to $\tilde{f}(O)$; then we can show the following result. 
\begin{lemma}\label{lemma:jamie}
Under Assumptions 1-3,
\begin{align*}
\gamma(\tilde{L}_0)&=\tilde{\mathbb{E}}\{Y_1(1)-Y_1(0)\}|\tilde{L}_0,A=1\}
\end{align*}
with probability one. Further, under Assumption 1, \begin{align*}
\tilde{\mathbb{E}}\{Y_1(0)-Y_0|\tilde{L}_0,A=1\}=\tilde{\mathbb{E}}\{Y_1(0)-Y_0|\tilde{L}_0,A=0\}
\end{align*}
with probability one.
\end{lemma}
\begin{proof}
In the following, we will use that both $\tilde{f}(X_0|\tilde{L}_0)=f(X_0|A=1,\tilde{L}_0)$ (by choice), and also $\tilde{f}(X_0|A=1,\tilde{L}_0)=f(X_0|A=1,\tilde{L}_0)$; the latter holds because
\begin{align*}
\tilde{f}(X_0|A=1,\tilde{L}_0)\tilde{f}(A=1|\tilde{L}_0)&=\tilde{f}(A=1|X_0,\tilde{L}_0)\tilde{f}(X_0|\tilde{L}_0),\\
\implies \tilde{f}(X_0|A=1,\tilde{L}_0)\tilde{f}(A=1|\tilde{L}_0)&=\tilde{f}(A=1|\tilde{L}_0)\tilde{f}(X_0|\tilde{L}_0),&\textrm{($A \perp X_0|L_0$ under  $\tilde{f}(O)$)}\\ 
\implies \tilde{f}(X_0|A=1,\tilde{L}_0)&=\tilde{f}(X_0|\tilde{L}_0),\\
\implies \tilde{f}(X_0|A=1,\tilde{L}_0)&=f(X_0|A=1,L_0)
\end{align*}
with probability one. Therefore $\tilde{f}(X_0|\tilde{L}_0)=\tilde{f}(X_0|A=1,\tilde{L}_0)$.

Then we have
\begin{align*}
\mathbb{E}\{Y_1(1)|\tilde{L}_0,A=1\}=&\mathbb{E}(Y_1|\tilde{L}_0,A=1) &\textrm{(Assumption 2)}\\
&=\mathbb{E}\{\mathbb{E}(Y_1|L_0,A=1)|\tilde{L}_0,A=1\}\\
&=\mathbb{E}\{\tilde{\mathbb{E}}(Y_1|L_0,A=1)|\tilde{L}_0,A=1\} &(f(Y_1,Y_0|L_0,A)=\tilde{f}(Y_1,Y_0|L_0,A))\\
&=\tilde{\mathbb{E}}\{\tilde{\mathbb{E}}(Y_1|L_0,A=1)|\tilde{L}_0,A=1\} &(f(X_0|A=1,\tilde{L}_0)=\tilde{f}(X_0|A=1,L_0))\\
&=\tilde{\mathbb{E}}(Y_1|\tilde{L}_0,A=1)\\
&=\tilde{\mathbb{E}}\{Y_1(1)|\tilde{L}_0,A=1\}  &\textrm{(Assumption 2)}
\end{align*}
with probability one. 

Next, we first note that  Assumption 1 implies 
\begin{align*}
\tilde{\mathbb{E}}(Y_1-Y_0|L_0,A=0)=\tilde{\mathbb{E}}\{Y_1(1)-Y_0|L_0,A=1\}
\end{align*}
since $\tilde{f}(O)$ and $f(O)$ do not differ in terms of the conditional density of $Y_1$ and $Y_0$ given $A$ and $L_0$. Further,
\begin{align*}
&\mathbb{E}\{Y_1(0)|\tilde{L}_0,A=1\}\\
&=\mathbb{E}[\mathbb{E}\{Y_1(0)|L_0,A=0\}\\&\quad+\mathbb{E}\{Y_1(0)|L_0,A=1\}-\mathbb{E}\{Y_1(0)|L_0,A=0\}|\tilde{L}_0,A=1]\\
&=\mathbb{E}[\mathbb{E}\{Y_1(0)|L_0,A=0\}+\mathbb{E}(Y_0|L_0,A=1)-\mathbb{E}(Y_0|L_0,A=0)|\tilde{L}_0,A=1] &\textrm{(Assumption 1)}\\
&=\mathbb{E}\{\mathbb{E}(Y_1-Y_0|L_0,A=0)+\mathbb{E}(Y_0|L_0,A=1)|\tilde{L}_0,A=1\} &\textrm{(Assumption 2)}\\
&=\mathbb{E}\{\tilde{\mathbb{E}}(Y_1-Y_0|L_0,A=0)+\tilde{\mathbb{E}}(Y_0|L_0,A=1)|\tilde{L}_0,A=1\} &(f(Y_1,Y_0|L_0,A)=\tilde{f}(Y_1,Y_0|L_0,A))\\
&=\mathbb{E}[\tilde{\mathbb{E}}\{Y_1(0)|L_0,A=1\}|\tilde{L}_0,A=1]\\
&=\tilde{\mathbb{E}}[\tilde{\mathbb{E}}\{Y_1(0)|L_0,A=1\}|\tilde{L}_0,A=1] &(f(X_0|A=1,L)=\tilde{f}(X_0|A=1,L))
\end{align*}
with probability one, where in the penultimate equality we invoke Assumptions 1 and 2 under $\tilde{f}(O)$ rather than $f(O)$.

To show the second part of the Lemma, 
\begin{align*}
&\tilde{\mathbb{E}}\{Y_1(0)-Y_0|\tilde{L}_0,A=1\}\\&=\tilde{\mathbb{E}}[\tilde{\mathbb{E}}\{Y_1(0)-Y_0|L_0,A=1\}|\tilde{L}_0,A=1]\\
&=\tilde{\mathbb{E}}[\tilde{\mathbb{E}}\{Y_1(0)-Y_0|L_0\}|\tilde{L}_0,A=1] &\textrm{(Assumption 1)}\\
&=\tilde{\mathbb{E}}[\tilde{\mathbb{E}}\{Y_1(0)-Y_0|L_0\}|\tilde{L}_0] &(\tilde{f}(X_0|A=1,\tilde{L}_0)=\tilde{f}(X_0|\tilde{L}_0))
\end{align*}
with probability one, where we note that Assumption 1 can be applied both under $f(O)$ and $\tilde{f}(O)$, since it is conditional on $L_0$ and $A$.
\end{proof}
The first part of this lemma gives a representation of the target parameter under $\tilde{f}(O)$. The second part shows that parallel trends holds conditional on $\tilde{L}_0$ under the law $\tilde{f}(O)$, which would not be true under $f(O)$.  Recall that under $f(O)$, our target parameter is an average of the conditional treatment effect on the treated, where we marginalize over variables $X_0$ which are necessary for identification but which does not define the contrast of interest. Since under $\tilde{f}(O)$, parallel trends holds conditional on $\tilde{L}_0$, Lemma \ref{lemma:jamie} implies that the marginal contrast under $f(O)$ is equal to the non-marginal effect under $\tilde{f}(O)$. This has the following important corollary:
\begin{corollary}\label{corollary:jamie}
Under the conditions of Lemma \ref{lemma:jamie}
\begin{align*}
\tilde{\mathbb{E}}\{Y_1-Y_0-\gamma(\tilde{L}_0)A|\tilde{L}_0,A\}=\tilde{\mathbb{E}}\{Y_1-Y_0-\gamma(\tilde{L}_0)A|\tilde{L}_0\}
\end{align*}
with probability one. 
\end{corollary}

\begin{proof}
\begin{align*}
&\tilde{\mathbb{E}}\{Y_1-Y_0-\gamma(\tilde{L}_0)A|\tilde{L}_0,A\}\\
&=\tilde{\mathbb{E}}(Y_1-Y_0|\tilde{L}_0,A)-\tilde{\mathbb{E}}\{Y_1(A)-Y_1(0)|\tilde{L}_0,A\} &\textrm{(Lemma \ref{lemma:jamie})}\\
&=\tilde{\mathbb{E}}\{Y_1(0)-Y_0|\tilde{L}_0,A\} &\textrm{(Assumption 1)}\\
&=\tilde{\mathbb{E}}\{Y_1(0)-Y_0|\tilde{L}_0\} &\textrm{(Lemma \ref{lemma:jamie})}
\end{align*}
with probability one. Since the right hand side of the final expectation does not depend on $A$, the result follows. 
\end{proof}

The conditional moment restriction above implies that under $\tilde{f}(O)$, we can use methods developed for \textit{semiparametric regression} for estimating $\psi^*$.

Specifically, we could use the doubly robust $g$-estimating functions in \citet{us2022}
\[q^\ddagger(\tilde{L}_0)\{A-\tilde{\mathbb{E}}(A|\tilde{L}_0)\}[Y_1-Y_0-A\gamma(\tilde{L}_0;\psi^*)-\tilde{\mathbb{E}}\{Y_1-Y_0-A\gamma(\tilde{L}_0;\psi^*)|\tilde{L}_0\}],\]
 which were developed for the non-marginal effect in the treated, to infer $\psi^*$; here, $q^\ddagger(\tilde{L}_0)$ is an arbitrary function of $\tilde{L}_0$. 

However, we do not have access to data generated under $\tilde{f}(O)$, but only the true law $f(O)$. Therefore, we will have to re-weight our data via the likelihood ratio:
\begin{align*}
\frac{\tilde{f}(X_0|\tilde{L}_0)}{f(X_0|\tilde{L}_0)}\frac{\tilde{f}(A|\tilde{L}_0)}{f(A|L_0)}=\frac{f(X_0|\tilde{L}_0,A=1)}{f(X_0|\tilde{L}_0)}\frac{\tilde{f}(A|\tilde{L}_0)}{f(A|L_0)}=\frac{\pi^*(L_0)}{f(A|\tilde{L}_0)}\frac{\tilde{f}(A|\tilde{L}_0)}{f(A|L_0)}.
\end{align*}
where in the last equality we used Bayes' formula.  If the law $f(A|L_0)$ was known, then $\psi^*$ could be estimated via the weighted estimating functions $\pi^*(L_0)\tilde{\phi}(O,\psi^*,q^\ddagger)$, where 
\[\tilde{\phi}(O,\psi^*,q^\ddagger)\equiv q^\ddagger(\tilde{L}_0)\frac{\tilde{f}(A|\tilde{L}_0)}{f(A=1|\tilde{L}_0)f(A|L_0)}\{A-\tilde{\mathbb{E}}(A|\tilde{L}_0)\}[Y_1-Y_0-A\gamma(\tilde{L}_0;\psi^*)-\tilde{\mathbb{E}}\{Y_1-Y_0-A\gamma(\tilde{L}_0;\psi^*)|\tilde{L}_0\}].\]
For fixed $q^\ddagger$, one can further improve efficiency by subtracting  from $\pi^*(L_0)\tilde{\phi}(O,\psi^*,q^\ddagger)$ its projection onto the scores for $f(A|L_0)$:
\begin{align}\label{equation:proj_sub}
\pi^*(L_0)\tilde{\phi}(O,\psi^*,q^\ddagger)-\pi^*(L_0)[\mathbb{E}\{\tilde{\phi}(O,\psi^*,q^\ddagger)|A,L_0\}-\mathbb{E}\{\tilde{\phi}(O,\psi^*,q^\ddagger)|L_0\}],
\end{align}
following Theorem 3.2 of \citet{robins2000marginal}. However, this will not deliver an (unscaled) influence function in the set given in Theorem \ref{theorem:pe_IFs}. This is because the propensity score is not ancillary for marginal treatment effect on the treated functionals, meaning that more efficient estimators can be obtained when the law $f(A|L_0)$ is known \citep{hahn1998role}.

In order to recover the set in Theorem \ref{theorem:pe_IFs},  noting that $\pi^*(L_0)$ is a conditional expectation, one can apply standard results to show that if we fluctuate only $\pi^*(L_0)$, the pathwise derivative for $\mathbb{E}\{\pi^*(L_0)\tilde{\phi}(O,\psi^*,q^\ddagger)\}$ is 
\[\{A-\pi^*(L_0)\}\mathbb{E}\{\tilde{\phi}(O,\psi^*,q^\ddagger)|L_0\}\]
since $\pi^*(L_0)$ is a conditional expectation. It therefore follows that the (unscaled) influence functions under model $\mathcal{M}$ can be obtained as 
\begin{align*}
&\pi^*(L_0)\tilde{\phi}(O,\psi^*,q^\ddagger)-\pi^*(L_0)[\mathbb{E}\{\tilde{\phi}(O,\psi^*,q^\ddagger)|A,L_0\}-\mathbb{E}\{\tilde{\phi}(O,\psi^*,q^\ddagger)|L_0\}]+\{A-\pi^*(L_0)\}\mathbb{E}\{\tilde{\phi}(O,\psi^*,q^\ddagger)|L_0\}\\
&=\pi^*(L_0)[\tilde{\phi}(O,\psi^*,q^\ddagger)-\mathbb{E}\{\tilde{\phi}(O,\psi^*,q^\ddagger)|A,L_0\}]+A\mathbb{E}\{\tilde{\phi}(O,\psi^*,q^\ddagger)|L_0\}.
\end{align*}
which recovers the result in Theorem \ref{theorem:pe_IFs}. In the time-varying setting considered in the remainder of the paper, an analogous connection can be made.
\section{Supplementary lemmas}

\begin{lemma}\label{lemma:cf}
Suppose that the conditions of Theorem \ref{estimation_theorem_2t} hold. Then we have the following asymptotic expansion of the cross-fit estimator $\hat{\psi}^{cf}$:
\begin{align*}
\sqrt{N}(\hat{\psi}^{cf}-\psi^{*})=&D(\psi^*,q)^{-1}\sqrt{N}(\mathbb{P}_N-\mathbb{P})\left\{\sum_{m=0}^K\phi_m(\psi^*,q_m,\pi^*,\mu^*)\right\}\\&+\sqrt{N}O_\mathbb{P}\left(\sum^{S}_{s=1}\left(\frac{n_s}{N}\right)\mathbb{P}\left\{\sum_{m=0}^K \phi_m(\psi^*,q_m,\hat{\pi}^{(-s)},\hat{\mu}^{(-s)})\right\}\right)+o_\mathbb{P}\left(1\right).
\end{align*}
\end{lemma}

\begin{proof}
Recall that $\hat{\psi}^{cf}=\sum^{S}_{s=1}\left(\frac{n_s}{N}\right)\hat{\psi}^{(s)}$. Under the conditions of Theorem \ref{estimation_theorem_2t} and applying Lemma 3 in \citet{kennedy2023semiparametric}, we have that 
\begin{align*}
\hat{\psi^{(s)}}-\psi^*=&D(\psi^*,q)^{-1}(\mathbb{P}_{n_s}-\mathbb{P})\left\{\sum_{m=0}^K\phi_m(\psi^*,q_m,\pi^*,\mu^*)\right\}\\
&+O_\mathbb{P}\left(\mathbb{P}\left\{\sum_{m=0}^K \phi_m(O;\psi^*,q_m,\hat{\pi}^{(-s)},\hat{\mu}^{(-s)})\right\}\right)+o_\mathbb{P}(n_s^{-1/2})
\end{align*}
for $s=1,...,S$. Therefore
\begin{align*}
\hat{\psi}^{(cf)}=&\sum^S_{s=1}\left(\frac{n_s}{N}\right)\bigg[\psi^*+D(\psi^*,q)^{-1}(\mathbb{P}_{n_s}-\mathbb{P})\left\{\sum_{m=0}^K\phi_m(\psi^*,q_m,\pi^*,\mu^*)\right\}\\
&+O_\mathbb{P}\left(\mathbb{P}\left\{\sum_{m=0}^K \phi_m(\psi^*,q_m,\hat{\pi}^{(-s)},\hat{\mu}^{(-s)})\right\}\right)+o_\mathbb{P}(n_s^{-1/2})\bigg]\\
=&\psi^*+D(\psi^*,q)^{-1}(\mathbb{P}_{N}-\mathbb{P})\left\{\sum_{m=0}^K\phi_m(\psi^*,q_m,\pi^*,\mu^*)\right\}\\
&+O_\mathbb{P}\left( \sum^S_{s=1}\left(\frac{n_s}{N}\right)\mathbb{P}\left\{\sum_{m=0}^K \phi_m(\psi^*,q_m,\hat{\pi}^{(-s)},\hat{\mu}^{(-s)})\right\}\right)+o_\mathbb{P}(N^{-1/2})\bigg]\
\end{align*}
Then the result follows via re-arrangement of the above and multiplying both sides of the equality by $\sqrt{N}$.
\end{proof}

In the following lemma and proof, to simplify the exposition, we will omit in this proof the dependence of the nuisance parameters on covariate and treatment histories.
\begin{lemma}\label{lemma:drift}
For all $m=0,...,K$ and $k=m+1,...,K+1$, under the mSNMM we have
\begin{align*}
&\mathbb{E}\{\epsilon_{mk}(O;\psi^*,\pi^\dagger,\mu^\dagger)|\bar{L}_m,\bar{A}_{m-1}\}\\&=\frac{(\pi_m^*-\pi_m^\dagger)}{(1-\pi_m^\dagger)}(\mu^*_{m,k}-\mu^\dagger_{m,k}) \\&\quad+\mathds{I}_{mk} \mathbb{E} \left[\sum^{k-1}_{j=m+1} \left\{ \prod^{j-1}_{r=m}\frac{(1-A_r)}{(1-\pi^\dagger_r)}\right\}\frac{(\pi^*_{j}-\pi^\dagger_{j})}{(1-\pi^\dagger_{j})}(\mu^*_{jk}-\mu^\dagger_{jk})|\bar{L}_m,\bar{A}_{m-1}\right]
\end{align*}
where $\pi_m^\dagger(\bar{L}_m,\bar{A}_{m-1)}$ and $\mu_{mk}^\dagger(\bar{L}_m,\bar{A}_{m-1})$ are arbitrary functions subject only to the constraint that $\{1-\pi_m^\dagger(\bar{L}_m,\bar{A}_{m-1})\}$ is bounded away from zero with probability one.
\end{lemma}

\begin{proof}
We note for all $m=0,...,K$ and $k=m+1,...,K+1$ we have 
\begin{align*}
&\mathbb{E}\{\epsilon_{mk}(O;\psi^*,\pi^\dagger,\mu^\dagger)|\bar{L}_m,\bar{A}_{m-1}\}&\\&=\mathbb{E} \bigg[\dot{U}_{mk}(\psi^*,\mu^\dagger_{mk}) -\sum^{k-1}_{j=m} \left\{ \prod^{j}_{r=m}\frac{(1-A_r)}{(1-\pi^\dagger_r)}\right\}(\mu^\dagger_{j+1,k}-\mu^\dagger_{jk})|\bar{L}_m,\bar{A}_{m-1}\bigg]& \textrm{(By definition)}\\
&=(\mu^*_{m,k}-\mu^\dagger_{m,k})-\mathbb{E}\left[\sum^{k-1}_{j=m} \left\{ \prod^{j}_{r=m}\frac{(1-A_r)}{(1-\pi^\dagger_r)}\right\}(\mu^\dagger_{j+1,k}-\mu^\dagger_{jk})|\bar{L}_m,\bar{A}_{m-1}\right]& \textrm{(mSNMM restriction)}
\end{align*}
which we will make use of frequently in the remainder of the proof. 
    
We will proceed using a reverse induction argument. First, choose $m=K$. Then 
\begin{align*}
&\mathbb{E}\{\epsilon_{K,K+1}(O;\psi^*,\pi^\dagger,\mu^\dagger)|\bar{L}_{K}
,\bar{A}_{K-1}\}\\
&=(\mu^*_{K,K+1}-\mu^\dagger_{K,K+1})-\mathbb{E}\left[\frac{(1-A_{K})}{(1-\pi_K^\dagger)}(Y_{K+1}-Y_K-\mu^\dagger_{K,K+1})|\bar{L}_{K}
,\bar{A}_{K-1}\right]\\
&=\frac{(\pi_{K}^*-\pi_{K}^\dagger)}{(1-\pi_{K}^\dagger)}(\mu^*_{K,K+1}-\mu^\dagger_{K,K+1})
\end{align*}
using the law of iterated expectation. Moreover, for  $m=K-1$, 
\begin{align*}
&\mathbb{E}\{\epsilon_{K-1,K+1}(O;\psi^*,\pi^\dagger,\mu^\dagger)|\bar{L}_{K-1}
,\bar{A}_{K-2}\}\\
&=(\mu^*_{K-1,K+1}-\mu^\dagger_{K-1,K+1})-\mathbb{E}\left[\frac{(1-A_{K-1})}{(1-\pi_{K-1}^\dagger)}(\mu^\dagger_{K,K+1}-\mu^\dagger_{K-1,K+1})|\bar{L}_{K-1}
,\bar{A}_{K-2}\right]\\
&\quad -\mathbb{E}\left[\frac{(1-A_{K-1})}{(1-\pi_{K-1}^\dagger)}\frac{(1-A_{K})}{(1-\pi_{K}^\dagger)}(Y_{K+1}-Y_K-\mu^\dagger_{K,K+1})|\bar{L}_{K-1}
,\bar{A}_{K-2}\right]\\
&=(\mu^*_{K-1,K+1}-\mu^\dagger_{K-1,K+1})\\
&\quad -\mathbb{E}\left(\frac{(1-A_{K-1})}{(1-\pi_{K-1}^\dagger)}\left[(\mu^*_{K,K+1}-\mu^\dagger_{K-1,K+1})-\mathbb{E}\{\epsilon_{K,K+1}(O;\psi^*,\pi^\dagger,\mu^\dagger)|\bar{L}_{K}
,\bar{A}_{K-1}\}\right]|\bar{L}_{K-1}
,\bar{A}_{K-2}\right)\\
&=(\mu^*_{K-1,K+1}-\mu^\dagger_{K-1,K+1})\\
&\quad -\mathbb{E}\left(\frac{(1-A_{K-1})}{(1-\pi_{K-1}^\dagger)}\left\{(\mu^*_{K,K+1}-\mu^\dagger_{K-1,K+1})-\frac{(\pi_{K}^*-\pi_{K}^\dagger)}{(1-\pi_{K}^\dagger)}(\mu^*_{K,K+1}-\mu^\dagger_{K,K+1})
\right\}|\bar{L}_{K-1}
,\bar{A}_{K-2}\right)\\
&=\frac{(\pi_{K-1}^*-\pi_{K-1}^\dagger)}{(1-\pi_{K-1}^\dagger)}(\mu^*_{K-1,K+1}-\mu^\dagger_{K-1,K+1})\\
&\quad +\mathbb{E}\left\{\frac{(1-A_{K-1})}{(1-\pi_{K-1}^\dagger)}\frac{(\pi_{K}^*-\pi_{K}^\dagger)}{(1-\pi_{K}^\dagger)}(\mu^*_{K,K+1}-\mu^\dagger_{K,K+1})
|\bar{L}_{K-1}
,\bar{A}_{K-2}\right\}
\end{align*}
where the third equality uses the result for $m=K$ and the fourth follows by the law of iterated expectation.

Next, for any $0\leq m<K-1$, we will assume that 
\begin{align}\label{equation:ind_hyp}
&\mathbb{E}\{\epsilon_{{m+1},K+1}(O;\psi^*,\pi^\dagger,\mu^\dagger)|\bar{L}_{m+1},\bar{A}_{m}\}\nonumber\\&=\frac{(\pi_{m+1}^*-\pi_{m+1}^\dagger)}{(1-\pi_{m+1}^\dagger)}(\mu^*_{{m+1},K+1}-\mu^\dagger_{{m+1},K+1})\nonumber\\&\quad+ \mathbb{E} \left[\sum^{K}_{j=m+2} \left\{ \prod^{j-1}_{r=m+1}\frac{(1-A_r)}{(1-\pi^\dagger_r)}\right\}\frac{(\pi^*_{j}-\pi^\dagger_{j})}{(1-\pi^\dagger_{j})}(\mu^*_{j,K+1}-\mu^\dagger_{j,K+1})|\bar{L}_{m+1},\bar{A}_{m}\right]
\end{align}
Then 
\begin{align*}
&\mathbb{E}\{\epsilon_{m,K+1}(O;\psi^*,\pi^\dagger,\mu^\dagger)|\bar{L}_m,\bar{A}_{m-1}\}\\
&=(\mu^*_{m,K+1}-\mu^\dagger_{m,K+1}) - \mathbb{E}\left[\sum^{K}_{j=m} \left\{ \prod^{j}_{r=m}\frac{(1-A_r)}{(1-\pi^\dagger_r)}\right\}(\mu^\dagger_{j+1,K+1}-\mu^\dagger_{j,K+1})|\bar{L}_m,\bar{A}_{m-1}\right]\\
&=(\mu^*_{m,K+1}-\mu^\dagger_{m,K+1}) - \mathbb{E}\bigg[\frac{(1-A_m)}{(1-\pi^\dagger_m)}(\mu^\dagger_{m+1,K+1}-\mu^\dagger_{m,K+1})+ \sum^{K}_{j=m+1} \left\{ \prod^{j}_{r=m}\frac{(1-A_r)}{(1-\pi^\dagger_r)}\right\}(\mu^\dagger_{j+1,K+1}-\mu^\dagger_{j,K+1})|\bar{L}_m,\bar{A}_{m-1}\bigg]\\
&=(\mu^*_{m,K+1}-\mu^\dagger_{m,K+1})- \mathbb{E}\left(\frac{(1-A_m)}{(1-\pi^\dagger_m)}[\mu^*_{m+1,K+1}-\mu^\dagger_{m,K+1}-\mathbb{E}\{\epsilon_{m+1,K+1}(O;\psi^*,\pi^\dagger,\mu^\dagger)|\bar{L}_{m+1},\bar{A}_{m}\}]|\bar{L}_m,\bar{A}_{m-1}\right)\\
&=\frac{(\pi_m^*-\pi_m^\dagger)}{(1-\pi_m^\dagger)}(\mu^*_{m,K+1}-\mu^\dagger_{m,K+1})+\mathbb{E}\left[\frac{(1-A_m)}{(1-\pi^\dagger_m)}\mathbb{E}\{\epsilon_{m+1,K+1}(O;\psi^*,\pi^\dagger,\mu^\dagger)|\bar{L}_{m+1},\bar{A}_{m}\}|\bar{L}_m,\bar{A}_{m-1}\right]\\
&=\frac{(\pi_m^*-\pi_m^\dagger)}{(1-\pi_m^\dagger)}(\mu^*_{m,K+1}-\mu^\dagger_{m,K+1})+\mathbb{E}\left[\frac{(1-A_m)}{(1-\pi^\dagger_m)}\frac{(\pi_{m+1}^*-\pi_{m+1}^\dagger)}{(1-\pi_{m+1}^\dagger)}(\mu^*_{{m+1},K+1}-\mu^\dagger_{{m+1},K+1})|\bar{L}_m,\bar{A}_{m-1}\right]\\
&\quad+\mathbb{E}\left(\frac{(1-A_m)}{(1-\pi^\dagger_m)}\mathbb{E} \left[\sum^{K}_{j=m+2} \left\{ \prod^{j-1}_{r=m+1}\frac{(1-A_r)}{(1-\pi^\dagger_r)}\right\}\frac{(\pi^*_{j}-\pi^\dagger_{j})}{(1-\pi^\dagger_{j})}(\mu^*_{j,K+1}-\mu^\dagger_{j,K+1})|\bar{L}_{m+1},\bar{A}_{m}\right]|\bar{L}_m,\bar{A}_{m-1}\right)\\
&=\frac{(\pi_m^*-\pi_m^\dagger)}{(1-\pi_m^\dagger)}(\mu^*_{m,K+1}-\mu^\dagger_{m,K+1})+\mathbb{E}\left[\sum^{K}_{j=m+1} \left\{ \prod^{j-1}_{r=m}\frac{(1-A_r)}{(1-\pi^\dagger_r)}\right\}\frac{(\pi^*_{j}-\pi^\dagger_{j})}{(1-\pi^\dagger_{j})}(\mu^*_{j,K+1}-\mu^\dagger_{j,K+1})|\bar{L}_m,\bar{A}_{m-1}\right]
\end{align*}
where the penultimate equality uses the induction hypothesis \eqref{equation:ind_hyp}. The result has then been shown for $k=K+1$, and the main result follows upon noting that an identical argument holds for arbitrary $0\leq k<K$.
\end{proof}

\begin{lemma}\label{lemma:product_rate}
Suppose that for $m=0,...,K$, $\{1-\hat{\pi}_m(\bar{L}_m,\bar{A}_{m-1})\}\geq c>0$ with probability one for a constant $c$. Then
\begin{align*}
&\sum^{K}_{m=0}\sum^{K+1}_{k=m+1}\mathbb{P}\left[\Bigg|\frac{(\pi_m^*-\hat{\pi}_m)}{(1-\hat{\pi}_m)}(\mu^*_{m,k}-\hat{\mu}_{m,k})+\mathds{I}_{mk}\sum^{k-1}_{j=m+1} \left\{ \prod^{j-1}_{r=m}\frac{(1-A_r)}{(1-\hat{\pi}_r)}\right\}\frac{(\pi^*_{j}-\hat{\pi}_{j})}{(1-\hat{\pi}_{j})}(\mu^*_{jk}-\hat{\mu}_{jk})\Bigg|\right]\\&=O_{\mathbb{P}}\left(\sum_{m=0}^{K} \sum_{k=m+1}^{K+1}\norm{\hat{\pi}_{m} - \pi_{m}^*}\norm{\hat{\mu}_{mk}-\mu_{mk}^*}\right)
\end{align*}
\end{lemma}

\begin{proof}
For an arbitrary $m$ and $k$, we have 
\begin{align*}
&\mathbb{P}\left[\Bigg|\frac{(\pi_m^*-\hat{\pi}_m)}{(1-\hat{\pi}_m)}(\mu^*_{m,k}-\hat{\mu}_{m,k})+\mathds{I}_{mk}\sum^{k-1}_{j=m+1} \left\{ \prod^{j-1}_{r=m}\frac{(1-A_r)}{(1-\hat{\pi}_r)}\right\}\frac{(\pi^*_{j}-\hat{\pi}_{j})}{(1-\hat{\pi}_{j})}(\mu^*_{jk}-\hat{\mu}_{jk})\Bigg|\right]\\
&\leq \mathbb{P}\bigg\{\bigg|\frac{(\pi_m^*-\hat{\pi}_m)}{(1-\hat{\pi}_m)}(\mu^*_{mk}-\hat{\mu}_{mk})\bigg|\bigg\}+\mathds{I}_{mk}\sum^{k-1}_{j=m+1} \mathbb{P}\bigg\{\bigg|\left\{\prod^{j-1}_{r=m}\frac{(1-A_r)}{(1-\hat{\pi}_r)}\right\}\frac{(\pi^*_{j}-\hat{\pi}_{j})}{(1-\hat{\pi}_{j})}(\mu^*_{jk}-\hat{\mu}_{jk})\bigg|\bigg\}\\
&=\mathbb{P}\left(\frac{|\pi_m^*-\hat{\pi}_m|}{|1-\hat{\pi}_m|}|\mu^*_{mk}-\hat{\mu}_{mk}|\right)+\mathds{I}_{mk}\sum^{k-1}_{j=m+1} \mathbb{P}\left(\left\{\prod^{j-1}_{r=m}\frac{|1-A_r|}{|1-\hat{\pi}_r|}\right\}\frac{|\pi^*_{j}-\hat{\pi}_{j}|}{|1-\hat{\pi}_{j}|}|\mu^*_{jk}-\hat{\mu}_{jk}|\right)\\
&\lesssim  \sum^{k-1}_{j=m} \norm{\hat{\pi}_{j} - \pi_{j}^*}\norm{\hat{\mu}_{jk}-\mu_{jk}^*}.
\end{align*}
where the first step follows from application of the triangle inequality, the third by assumption and the Cauchy-Schwarz inequality.

Summing across $m$ and $k$, this gives us 
\begin{align*}
&\sum^{K}_{m=0}\sum^{K+1}_{k=m+1}\mathbb{P}\left[\Bigg|\frac{(\pi_m^*-\hat{\pi}_m)}{(1-\hat{\pi}_m)}(\mu^*_{m,k}-\hat{\mu}_{m,k})+\mathds{I}_{mk}\sum^{k-1}_{j=m+1} \left\{ \prod^{j-1}_{r=m}\frac{(1-A_r)}{(1-\hat{\pi}_r)}\right\}\frac{(\pi^*_{j}-\hat{\pi}_{j})}{(1-\hat{\pi}_{j})}(\mu^*_{jk}-\hat{\mu}_{jk})\Bigg|\right]\\
&=O_{\mathbb{P}}\left(\sum_{m=0}^{K} \sum_{k=m+1}^{K+1} \sum_{j=m}^{k-1} \norm{\hat{\pi}_{j} - \pi_{j}^*}\norm{\hat{\mu}_{jk}-\mu_{jk}^*}\right)\\
&=O_{\mathbb{P}}\left(\sum_{m=0}^{K} \sum_{k=m+1}^{K+1}(m+1)\norm{\hat{\pi}_{m} - \pi_{m}^*}\norm{\hat{\mu}_{mk}-\mu_{mk}^*}\right)\\
&=O_{\mathbb{P}}\left(\sum_{m=0}^{K} \sum_{k=m+1}^{K+1}\norm{\hat{\pi}_{m} - \pi_{m}^*}\norm{\hat{\mu}_{mk}-\mu_{mk}^*}\right)
\end{align*}
since it is assumed that $K$ is fixed. 
\end{proof}

\end{document}